\documentclass[10pt,journal,comsoc]{IEEEtran}
\usepackage{graphicx}
\usepackage{multirow}
\usepackage{cite}
\usepackage{hyperref}
\ifCLASSINFOpdf
\else
\fi
\begin{document}
%
\title{A Survey on Open-Source-Defined Wireless Networks: Framework, Key Technology, and Implementation}

 \author{~Liqiang Zhao, ~Muhammad Muhammad Bala, ~Wu Gang, ~Pan Chengkang, ~Yuan Yannan, ~Tian Zhigang, ~Yu-Chee Tseng, ~Chen Xiang, ~Bin Shen, ~and Chih-Lin I \IEEEmembership{~}\IEEEmembership{~}~~\IEEEmembership{~~}
 	\thanks{This work was supported in part by National Key R\&D Program of China (2019YFE0196400),  Joint Project of China Mobile Research Institute \& X-NET (R202111101112JZC04), National Natural Science Foundation of China (61771358, 61901317, 62071352), Fundamental Research Funds for the Central Universities (JB190104), Science and Technology Plan of Xi'An City (2019217014GXRC006CG007-GXYD6.1), Joint Education Project between China and Central-Eastern European Countries (202005) and the 111 Project (B08038).
 				
 	L. Zhao and M. M. Bala are with the State Key Laboratory of Integrated Service Networks, School of Telecommunications Engineering, Xidian University, Xi’an 710071, China (e-mail: muhammadyakasai@hotmail.com; lqzhao@mail.xidian.edu.cn;). 
 	
 	G. Wu is with National Key Laboratory of Science and Technology on Communications, University of Electronic Science and Technology of China (e-mail:wugang99@uestc.edu.cn). 
 	
 	P. Chengkang is with China Mobile Research Institute (e-mail:panchengkang@chinamobile.com).
 	
 	Y. Yannan is with Communication Research Institute Vivo Mobile Communication Co., Ltd. Beijing, China (e-mail: yannan.yuan@vivo.com).
 	
 	T. Zhigang is with Beijing National Research Center for Information Science and Technology Tsinghua University Beijing, China (e-mail: zgtian@tsinghua.edu.cn).
 	
 	Y. Tseng is with college of Artificial Intelligence National Yang Ming Chiao Tung University Taiwan (email: yctseng@cs.nctu.edu.tw). 
 	
 	C. Xiang is with School of Electronics and Information Technology, Sun Yat-sen University, Guangzhou, Guangdong Province, China, 510006 (chenxiang@mail.sysu.edu.cn).
 	
 	B. Shen is with Chongqing Municipal Lab of Mobile Communication Technology, Chongqing 400-065, P. R. China (e-mail: shenbin@cqupt.edu.cn).
 	
 	I. Chih-Lin is with China Mobile Research Institute Beijing, China (e-mail: icl@chinamobile.com).
 	}
 }


%


\maketitle

\begin{abstract}
The realization of open-source-defined wireless networks in the telecommunication domain is accomplished through the fifth-generation network (5G). In contrast to it predecessors (3G and 4G), the 5G network can support a wide variety of heterogeneous use cases with challenging requirements from both the Internet and Internet of Things (IoT). The future sixth-generation (6G) network will not only extend 5G capabilities but also innovate new functionalities to address emerging academic and engineering challenges. The research community has identified these challenges could be overcome by open-source-defined wireless networks, which is based on open-source software and hardware. In this survey, we present an overview of different aspects of open-source-defined wireless networks, comprising motivation, frameworks, key technologies, and implementation. We start by introducing the motivation and explore several frameworks with classification into three different categories: black-box, grey-box, and white-box. We review research efforts related to open-source-defined Core Network (CN), Radio Access Network (RAN), Multi-access Edge Computing (MEC), the capabilities of security threats, open-source hardware, and various implementations, including testbeds. The last but most important in this survey, lessons learned, future research direction, open research issues, pitfalls and limitations of existing surveys on open-source wireless networks are included to motivate and encourage future research.

\end{abstract}

%
\IEEEpeerreviewmaketitle

\section{Introduction}\label{ZZ}

\IEEEPARstart{T}{he fifth} generation (5G) network has been recognized as a key enabler to address the increase in data traffic of mobile Internet and Internet of Things (IoT), which enable a large number of smart devices to communicate with one another\cite{1}. This triggers various service requirements, for example, millisecond delay is required by Internet of vehicle (IoV) and ultra-high-definition video 4k/8k demand for Gbps transmission rate. However, the sixth generation (6G) network is envisioned to provide a higher bit rate and low latency to services and applications with constraints requirement \cite{2}. Furthermore, such a network that is expected to satisfy the requirement of the Internet and IoT will be difficult to implement. It is undeniable that conventional network architectural hardware and software are both integrated closely together and cannot withstand the diverse evolving scenarios and requirements metrics. This proposed the vision for open-source-defined wireless networks which enable the deployment of open-source software on general-purpose hardware and permit the scaling and customization of users' networks by mobile operators.

Open-source software evolution can be dated back in history from the computing industry with computer operating system free delivery as part of the hardware purchase in 1950s\cite{3}. In recent years, open-source software has dominated an end-to-end landscape of software from the cloud to the edge and terminal devices. For example, OpenStack, Linux, Unix, and Android were developed as open-source software with code visibility to learn, use and contribute to the technology. On the other hand, the sixth-generation network should address and improve the open capability of network infrastructure that should be built upon open-source software and hardware to fully address the requirement of open-source-defined wireless networks. Software-Defined Networking (SDN)\cite{003}, and Network Function Virtualization (NFV)\cite{108}, will serve as the technology enablers to support the implementation of open-source-defined wireless networks and to realize RAN slicing as well as Multi-access Edge Computing (MEC)\cite{036} for edge device applications with low latency requirement. This enables resources in close proximity to the IoT devices, e.g., URLLC.   
\begin{figure*}[t]
	\begin{center}
		\includegraphics[width=7in]{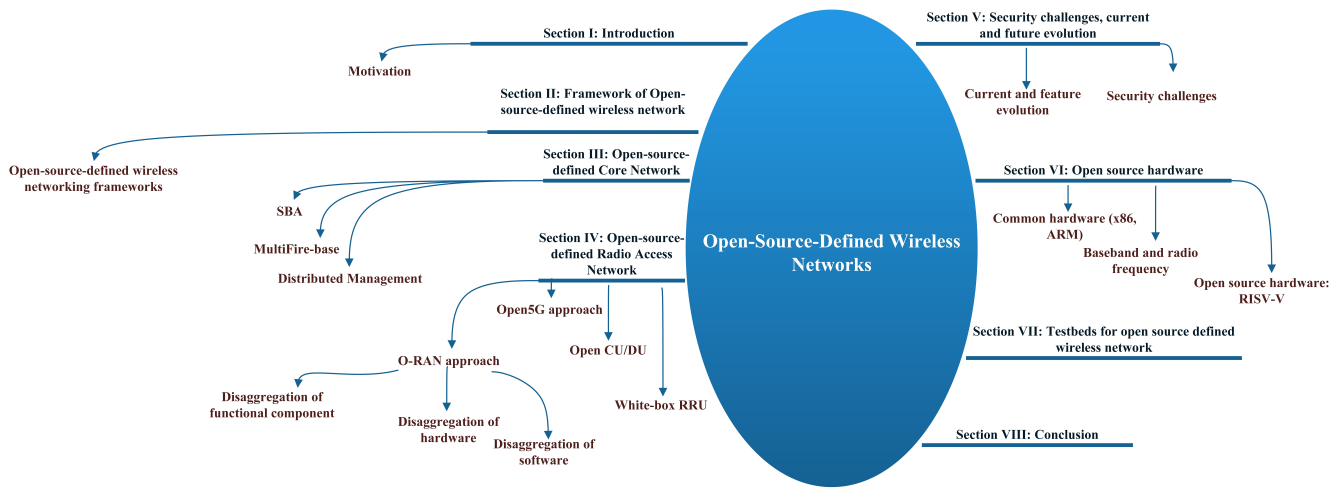}
		\caption{Structure of the survey}
		\label{fig1}
	\end{center}
\end{figure*}
This paper envisioned providing a survey of open-source-defined wireless networks focusing on different frameworks, categories, and key technology enablers in this novel research field. We also highlighted the key role of this research field in cellular networks for 5G and 6G.  
\subsection{Motivation and objective of open-source-defined wireless network} 
The concept of open network is a term used to describe network-based open standards and commodity hardware which enables compatibility in both hardware and software multi-vendor products. This provides the potential solution for the multi-vendor choice product. With complex infrastructure, close and proprietary interfaces which create vendor lock-in, and a centric cycle decade of single vendor maintenance with high CAPital EXpenditure (CAPEX) and OPerational EXpenditure (OPEX), that require good management, proper tools used for abstraction, and system organization. Openness is the key solution to the problem which includes both open-source community and standardization that encourage flexibility, interoperability and promote competition. 

However, most of the recent surveys and research articles on open-source wireless networks focus on the integration of open-source software with IoT devices, open-source testbeds, and lacks the detail implementation of open-source frameworks into the core network, RAN, and edge network. We cleave apart the ideology of centric perspective and give a broader view on the implementation of open-source-defined wireless networks to our readers. The principal idea of openness and expandability, the infrastructure which supports IoT permits 
to communicate with external applications and allows coordination among the particular component \cite{011}. This can be analyzed from a state-of-the-art idea with the SDN concept proposed in \cite{0012}. 
With the concept of an open network, the architecture allows many operators to use and share the same physical infrastructure. In the realization of this goal, the technology of SDN is used and CPRI over Ethernet is proposed in the fronthaul as the transport protocol, whereas in the backhaul the concept of distributed security with SDN is analyzed, which provides a low latency requirement. In the work of Idachaba \cite{015}, proposed architecture which relies on the collation strategies and centralized network planning including cloud based MSC, HLR/VLR database system to minimize operators' cost of infrastructural deployment.  

This network topology minimizes the deployment cost of a 5G network with an increase in network capacity to support the diverse requirement use cases proposed by the 5G network. The open network architecture relies on the collocation strategy, centralized network planning, and a cloud base mobile switching center (MSC) with a home/visitor location register (HLR/VLR) database system. In this concept, the MSC and the database service are provided by a third party, while the base station controller (BSC) and base transceiver station (BTS) is controlled by the individual operator’s own network domain. This open network architectural topology benefits different operators in a win-win fashion, such that traffic from one mobile operator can be routed through the infrastructure of another operator in absence of free channels to route traffic of required customers, sharing both of revenue by both operators. An open network can be analyzed from the three functional planes which are: the management plane, control plane, and data plane, with an open-source project making an impact on each plane and accelerating openness, flexibility, and interoperability with vendor-independent networking equipment. The control plane centralizes the logical decision-making, whereas the data plane forwards data traffic with instructions received on how to process and forward a packet, and the management plane defines the network policy. Moreover, this can be analyzed from open-source software, open-source project, and open API (i.e., application programmable interface). 
\subsubsection{Open-Source Software}
Open-source software (OSS) historic evolution dated back from first-generation computers in 1950 and 1960, and make a great impact in current wireless networks, e.g., it spanned the terminal with the android and the cloud with OpenStack. This software license enables the modification and distribution of source code. A very successful open-source project has been established with this ideology, e.g., Linux operating system. In recent years, OSS play a crucial role in the development of IoT applications, which differ from other non-IoT applications. For example, both mobile and web application development is different when compared to IoT application development. As such Corno et al. \cite{001} presented evidence on some open-source IoT software, that is characterized and considered by researchers with software engineering needs in the IoT domain. Due to challenges and requirements, expertise is needed, because of the diverse application features of IoT devices. The authors provided hints on the development of IoT software by analyzing the OSS project on GitHub repository through contributor's behavior, and the maturity of IoT software. RIOT \cite{012}, which is an open-source operating system design for low-end embedded devices in the IoT domain written from scratch with little memory requirement. This has made RIOT with characteristic features, such as code license and vendor-independent libraries making it unique and similar to Linux in an open-source ecosystem. 

OSS faces the problems of architectural degradation due to the persistent evolution of the system software with time. This implies the roles software architecture play in every system stage development, as such Baabad et al. \cite{002} present the main objective and a review in the architectural degradation of OSS with data collected from an online database (i.e., IEEE Explorer, Scopus, Web of Science, Wiley, ACM, Springer, ScienceDirect, and Taylor). The author's result shows that software evolution, lack of developer's awareness, and frequent changes are the prominent causes of software architectural degradation in the OSS project. The authors' proposed solution shows that the metric-based strategy addresses the problem, such as anomalies prioritization strategy, investigating and addressing architectural roles violation, architectural recovery strategy, and refactoring strategy. 
 
However, boosting the OSS community has made a large amount of code available that allows machine learning and deep learning technique to leverage this open-source code. With deep learning technique, motivates researchers and software engineers in learning and understand the vulnerability in open-source code projects using the deep neural technique \cite{006}.  Natella et al. \cite{007}, propose a methodology used in analyzing faults in OSS interface components which turns error at the components interface. This hinders experimental analysis on where, what, and when to inject interface error. The author's result shows that the conventional error model does not emulate accurately the software fault, rather require a rich interface error to be injected via both fail-stop behavior and data corruptions that focus on a large number of corrupted data structure. In general, the free open-source software community (FLOSS) contributes to the FLOSS project based on the onion model which identified core contributors who contribute to the major percentage of the source code, and secondary contributors/associate contribute as described in \cite{009}. The author's survey identifies the importance of encouraging contributors in the field of FLOSS projects with more than ten FLOSS communities investigating in the FLOSS context.

Integrating open-source software, SDN/NFV, and OpenStack into a full real-time stack, i.e., User Equipment (UE), and CN on an open-source hardware processor can lead to an efficient 5G/6G network. Mobile operators can now abstract the lower-level functionalities with flexibility in the deployment of mobile network functions, network programmable advantage\cite{0006}, network slicing, dynamic bandwidth customization to reduce delay, and assigning an optimal resource to a particular network demand \cite{6}. Telecom operators have figured out the way of managing their wireless network operation with software through virtualization \cite{0007}, e.g., the software can emulate and perform different requirement tasks of specialized hardware and can be upgraded and repaired remotely. This allows great control and ability to examine each line of code to fix errors and bugs, thereby increasing resiliency \cite{7}.
\subsubsection{Open-Source Project}
As presented in table \ref{tab.I}, many number of consortium projects are being carried out in progress of open-source-defined wireless networks, such as an open platform for network function virtualization (OPNFV) introduced by the European Telecommunication Standards Institute (ETSI) to support multiple SDN controllers, such as OpenDayLight, Open Network Operating System (ONOS), OpenContrail, which is based on OpenStack software with support for both x86 and ARM compatibility. Sionna is another open-source project based TensorFlow for link-level simulation on GPU-accelerated open-source library, which enable the prototyping of complex communication system architecture with high-level API support for the integration of neural networks. Sionna implement state-of-the-art algorithms used for bench-marking and end-to-end performance evaluation which permit effective research implementation of novel ideas. 
Sionna is developed for 6G physical layer research and support multi-user multiple input multiple output link-level simulation setups with 5G compliant codes \cite{0007x},\cite{0007y}.
\begin{table*}[]
	\caption{Summary of open-source effort and references}
	\label{tab.I}
	\centering
	\begin{tabular}{|c|c|c|c|c|}
		\hline
		\textbf{Networking area}                & \textbf{Reference design focus} & \textbf{Analysis}                                                                                                                                                 & \textbf{open-source effort}                                                                 & \textbf{Reference} \\ \hline
		\multirow{3}{*}{Infrastructure}         & Networking                      & \begin{tabular}[c]{@{}c@{}}Acceleration of packet \\ Processing technique\end{tabular}                                                                            & DPDK, VPP                                                                & \cite{9}                    \\ \cline{2-5} 
		& Hardware                        & \begin{tabular}[c]{@{}c@{}}High performance \\ and flexibility\end{tabular}                                                                                       & \begin{tabular}[c]{@{}c@{}}Open compute \\ project, O-RAN, P4\end{tabular}     & \cite{10}, \cite{15}, \cite{11}                   \\ \cline{2-5} 
		& Operating system                & \begin{tabular}[c]{@{}c@{}}Used of white box\\  in carrier grade network\end{tabular}                                                                             & \begin{tabular}[c]{@{}c@{}}Linux, Disaggregated \\ network \\ operating system\end{tabular} & \cite{12}, \cite{13}                   \\ \hline
		Access network                          & Radio                           & \begin{tabular}[c]{@{}c@{}}Implementation of \\ 5G radio access network\end{tabular}                                                                              & Openair5G, O-RAN                                                             & \cite{14}, \cite{15}                    \\ \hline
		Core network                            & Wireless core network           & \begin{tabular}[c]{@{}c@{}}Implementation of EPC \\ and \\ 5G NGC (next generation core)\end{tabular}                                        & \begin{tabular}[c]{@{}c@{}}openairCN, \\ M-CORD NGIC\end{tabular}                           & \cite{16}, \cite{17}                   \\ \hline
		\multirow{9}{*}{Management and control} & Networking                      & \begin{tabular}[c]{@{}c@{}}Process of packet \\ and flow control\end{tabular}                                                                                     & \begin{tabular}[c]{@{}c@{}}OpenDayLight, ONOS,\\ Open vSwitch, FD.io\end{tabular}           & \cite{18}, \cite{19}, \cite{20}, \cite{21}                   \\ \cline{2-5} 
		& Virtualization                  & \begin{tabular}[c]{@{}c@{}}The virtualization of compute \\ resources that is shared \\ among different application\\  and logical network\end{tabular}           & \begin{tabular}[c]{@{}c@{}}OpenStack, Kubernetes, \\ Docker\end{tabular}                    & \cite{22}, \cite{23}, \cite{24}                   \\ \cline{2-5} 
		& Automation                      & \begin{tabular}[c]{@{}c@{}}Providing management\\ and orchestration tools \\ used in control of network \\ component through \\ virtualization layer\end{tabular} & ONAP, xRAN, O-RAN, Ansible                                                                         &   \cite{25}, \cite{26}, \cite{15}, \cite{27}                \\ \cline{2-5} 
		& Orchestration                   & \begin{tabular}[c]{@{}c@{}}The deployment of network \\ policy and performance analysis\end{tabular}                                                              & \begin{tabular}[c]{@{}c@{}}XOS, \\ open-source MANO\end{tabular}                            &   \cite{28}, \cite{29}                \\ \cline{2-5} 
		& Modelling                       & \begin{tabular}[c]{@{}c@{}}Tools for modelling and \\ defining the state of network\\  services\end{tabular}                                                      & YANG, JuJu, YAML                                                               & \cite{30}, \cite{31}, \cite{32}                    \\ \cline{2-5} 
		& DevOps                          & \begin{tabular}[c]{@{}c@{}}The development of software \\ methodology in deployment of \\ workload into NFV environment\end{tabular}                              & \begin{tabular}[c]{@{}c@{}}Consul, puppet, \\ Chef, Jinkins\end{tabular}                    &  \cite{33}, \cite{34}, \cite{35}, \cite{36}                  \\ \cline{2-5} 
		& Analytics                       & \begin{tabular}[c]{@{}c@{}}The continuous streaming \\ of data for service monitoring\end{tabular}                                                                & \begin{tabular}[c]{@{}c@{}}Apache kafka, \\ Apache Spark\end{tabular}                       & \cite{27}, \cite{38}                   \\ \cline{2-5} 
		& AI                              & AI frame work used in network                                                                                                                                     & Automation                                                             & \cite{39}                  \\ \cline{2-5} 
		& Edge computing                  & open-source software for edge                                                                                                                                     & Computing                                                                   & \cite{40}                    \\ \hline
		Security                                & Cyber security                  & Security for virtual infrastructure                                                                                                                               & SHIELD                                                                                      &     \cite{41}              \\ \hline
	\end{tabular}
\end{table*}
O-RAN software community and Telecom Infra Project (TIP) both are rallying industry standards toward open and virtualized mobile networks, with cloud-base mobile infrastructure service providers can have great benefits from the diverse supply chain, flexibility, deployment, and design. OpenAirInterface (OAI) is another open-source project whose purpose is softwarization of mobile network functions ranging from the network access, e.g., OAI-RAN to the core mobile network, e.g., OAI-CN used in 4G mobile network prototyping. Mosaic5G \cite{8} is a complementary project to OAI whose goal is building 5G agile services that leverages SDN/NFV, and Multi-access Edge Computing (MEC) technology enabling service-based 5G vision. Open Compute Project (OCP) addresses the current challenges of computing and storage that face existing infrastructure technology from the surge of data traffic, and compute power requirement by creating open-source hardware with open specifications to ensure scalability, innovation, and energy-efficient hardware, e.g., AMD open 3.0.

Open-source projects face challenges in the integration with real SDN/NFV deployment, performance, reliability, automation, and dissemination, with mobile operator's interest in the open-source community, e.g., OPNFV, OpenStack, DPDK (Data Plane Development Kit), and so on to prevent vendor lock-in through the use of standardizing and interoperable solutions can be identified.
\begin{itemize}
	\item Deployment: This can be identified as one of the challenges faced by the project, with the recent continuous integration (CI) and code review used in the open-source project which is a standard development in practice, from different contributors' involvement with challenges when trying to contribute. For example, operators could face difficulty in the deployment of their virtual network function (VNF) when the open-source project solution is on the GitHub repository.
	
	\item Performance: As part of the challenge by the open-source project is performance, as not all specialized tasks can be performed easily on general-purpose hardware but required a practical knowledge of the hardware and software for a better result. This knowledge is required by skilled telecom operators in achieving seamless operation.
	
	\item Reliability: open-source solution project is expected to be reliable in testing and verification across all domains before integrating into the existing system. Due to the procedure, the developmental method of some software cannot be verified, and require a method in identifying key areas for the insertion of error handling for better tolerance. 
\end{itemize}

In the open-source project, test cases and code coverage are important factors. The former allows developers to study and test the code for perfect software quality, whereas the latter provides developers with information on what part of the source code is not tested and can be exposed to the source of bugs. However, implementing a large number of code coverage proves that code coverage test cases should not be overestimated \cite{008}. In addition, FloWatcher-DPDK is a novel open-source project solution based-DPDK that is high-speed and lightweight software traffic monitoring, which provides a fine-grain statistic at packet and flow level.  FloWatcher-DPDK provides researchers with flexibility and a state-of-the-art open-source tool in monitoring and performance capability at high-speed \cite{029}.
\subsubsection{Open Application Programming Interface (Open API)}
In demand for multi-vendor product interoperability and flexibility, the open-source community project delivers agile products with open interfaces and programmable APIs across all layers of the network. Open API with modularity to the southbound interface of SDN enables a firm to believe that the project to be delivered will be vendor-independent and promote network flexibility. Jang et al. \cite{040} present the significance of open API Artificial Intelligence (AI) vision from open-source software taking into consideration factors, domestic companies consider in using open API. Such factors support the development of open-source software with new technologies which include financial, periodical, and environmental factors, with the increase in the number of practical users. OpenAPIDocGen \cite{041} is an open-source project for API auto-generate documentation system, that describes the data sources and techniques used in generating the different sections of the document, which improve the developmental efficiency to ensure the correctness of API usage. In a nutshell, open API provides a third-party application, for innovation, and multi-vendor product interoperability.

\subsection{Open-Source-Defined Cellular Networks for 5G and 6G}
In this subsection, we discuss open-source-defined cellular networks from Core Network (CN), Radio Access Network (RAN), and terminal. In the CN, Service-Base Architecture (SBA) is defined by 3rd Generation Partnership Project (3GPP) to achieve flexibility via modular and loosely couple architecture. The SBA enable the 5GC based modularity for cloud native implementation\cite{00015}. Open-source-defined cellular SBA extend via the open optical transport backhaul, to open Centralize Unit (CU)/Distributed Unit (DU), and the edge cloud. However, open-source project that enable network slicing, network management and orchestration in both CN and RAN cannot be overemphasis. From the RAN architectural perspective, open-source software and open software project, as well as open-source hardware defined the RAN disaggregation based 3GPP with open interfaces for multi-vendor product interoperability for RAN real-time requirement, flexibility and virtualization. This is achieved through splitting of the higher layer protocol stack (i.e., PDCP, SDAP, and RRC) as discuss in Section \ref{AA}, specified by 3GPP. This has enabled the deployment of CU, DU, Radio Unit (RU) and RAN Intelligent Controller (RIC) at the edge of the network. It is well known, open-source-defined cellular networks is envision for the next generation wireless network, e.g., 6G, from the CN to the edge RAN in addressing end-to-end services requirement.  
\subsection{Related Literature}
Multiple surveys on open-source wireless networks have been conducted recently. However, most of these surveys and research articles focus on the RAN, Core network,   
open-source software, frameworks, programmable and virtualize networks but lacks the detail implementation of these open-source frameworks into core, access, and edge network. We will proceed to differentiate these surveys from our survey.
The survey in \cite{b019} provide detail implementation of various open-source frameworks into core, access, and edge network, but lacks detail security related articles to the readers for open-source-defined wireless networks to learned lessons from and prevent the risk of security breach. As represented in table \ref{tab.X}, the different area of focus of recent surveys and research articles related to open-source wireless networks. In this survey, we are not only focusing on open-source software but we also focus on open-source hardware because coupling the two will give a precise understanding of open-source-defined wireless networks to a novel researcher in this field. 
\begin{table*}[]
	\caption{Summary of important surveys and articles on the open-source wireless network}
	\label{tab.X}
	\centering
		\begin{tabular}{|c|c|c|}
			\hline
			Relevant challenge &
			Ref. &
			Main Contribution \\ \hline
			\begin{tabular}[c]{@{}c@{}}The deluge of software solutions \\ whose interoperability and \\ operations are unclear.\end{tabular} &
			\cite{b019} &
			\begin{tabular}[c]{@{}c@{}}The provision of critical requirements on the limitations \\ of the state-of-the-art frameworks, hardware, testbeds, \\ and feasible direction toward open and programmable networks.\end{tabular} \\ \hline
			\begin{tabular}[c]{@{}c@{}}Volunteer contribution to the FLOSS \\ project based on the onion model\end{tabular} &
			\cite{009} &
			\begin{tabular}[c]{@{}c@{}}An explorative qualitative survey of 13 FLOSS communities \\ was investigated and framework management by the volunteers.\end{tabular} \\ \hline
			\begin{tabular}[c]{@{}c@{}}Larger scale OpenFlow infrastructure\\ deployment to enable the research community \\ to experiment with novel ideas and application testing.\end{tabular} &
			\cite{90} &
			\begin{tabular}[c]{@{}c@{}}Performance measurement of OpenFlow networks by modeling \\ and experimentation\end{tabular} \\ \hline
			Vendor monopolies of networks products &
			\cite{v} &
			\begin{tabular}[c]{@{}c@{}}The study of the architectural transformation and its influence \\ on modern cloud computing.\end{tabular} \\ \hline
			\begin{tabular}[c]{@{}c@{}}The impeding communications encountered through \\ infected wireless networks affect security and privacy.\end{tabular} &
			\cite{147} &
			\begin{tabular}[c]{@{}c@{}}The 5G security model, network softwarization security, \\ physical layer security of the core, and enabling technologies.\end{tabular} \\ \hline
			\begin{tabular}[c]{@{}c@{}}The security features, security requirements \\ and security solution.\end{tabular} &
			\cite{148} &
			\begin{tabular}[c]{@{}c@{}}The network security architecture and security functionality\\ of 3GPP 5G networks.\end{tabular} \\ \hline
			\begin{tabular}[c]{@{}c@{}}The capability of meeting the 5G latency requirement \\ from the 5G cellular network requirement.\end{tabular} &
			\cite{x015} &
			\begin{tabular}[c]{@{}c@{}}The detailed technology to achieve low latency requirements \\ in three domains (RAN, Core, and caching).\end{tabular} \\ \hline
			\begin{tabular}[c]{@{}c@{}}The integration of edge computing and name data \\ networking latency and backbone network traffic \\ with the processing of large user data in real-time.\end{tabular} &
			\cite{011} &
			\begin{tabular}[c]{@{}c@{}}Three main tiers of an architectural framework consist of IoT \\ end devices, edge node applications, and cloud node applications.\end{tabular} \\ \hline
			\begin{tabular}[c]{@{}c@{}}Qualitative analysis of publicly available IoT \\ and non-IoT projects on Github.\end{tabular} &
			\cite{xx015} &
			\begin{tabular}[c]{@{}c@{}}Open-source IoT project developmental effort to satisfy \\ software engineering needs.\end{tabular} \\ \hline
			\begin{tabular}[c]{@{}c@{}}The complementary network layer ULL supported by \\ IEEE 802.1 time-sensitive networking for low \\ end-to-end latency.\end{tabular} &
			\cite{001} &
			\begin{tabular}[c]{@{}c@{}}The support of ULL in 5G networks takes into consideration \\ fronthaul, backhaul, and network management.\end{tabular} \\ \hline
			\begin{tabular}[c]{@{}c@{}}The throughput, energy efficiency, latency, and security \\ enhancement.\end{tabular} &
			\multicolumn{1}{l|}{\cite{001a}} &
			\begin{tabular}[c]{@{}c@{}}The analysis enhancement of major C-RAN \\ research focuses on the recent advancement\end{tabular} \\ \hline
			\begin{tabular}[c]{@{}c@{}}The limitations and unpredictable issues of RAN \\ virtualization\end{tabular} &
			\multicolumn{1}{l|}{\cite{xx016}} &
			\begin{tabular}[c]{@{}c@{}}Provide generic concept, limitations, and \\ evolution of the C-RAN\end{tabular} \\ \hline
			\begin{tabular}[c]{@{}c@{}}The role of an open-source project, such as\\ OpenDayLight, OPNFV, OpenStack, M-CORD, and \\ ONAP in the 6G network\end{tabular} &
			\multicolumn{1}{l|}{\cite{xx017}} &
			\begin{tabular}[c]{@{}c@{}}The optimization of 6G network through open-source \\ \end{tabular} \\ \hline
		\end{tabular}
\end{table*}

In \cite{b019} programmable networks with innovative ideas in making the network scalable stem the SDN architecture by abstracting the lower infrastructure. The authors provide the recent open-source software and frameworks for 5G cellular networks, with full end-to-end stack. Name data networking and cloud computing technology are consider as the future internet in \cite{009}. The author's framework consist of three main tier, i.e., the end devices comprising of IoT device, edge computing for edge node application, cloud computing for cloud edge application to reduced latency and backbone traffic. Ultra-Low Latency (ULL) support in 5G networks is survey in \cite{90} with focus on fronthaul, backhaul, and network management. The authors surveyed IEEE 802.1 time sensitive networking standards and IETF deterministic networking standards for network layer ULL support. OpenFlow-based architecture details is discuss in \cite{v} describing the challenges OpenFlow-based networks devices faced at large scale deployment. The evolution of programmable networks and SDN is presented in \cite{147} that illustrate the features of SDN successive emergence over the past decade. The proliferation of research achievement that enable the IEEE standardization effort of network programmability through OpenFlow. Security and privacy details on core and enabling technologies are discuss in \cite{148}. Security monitoring and management of 5G networks and security measures as well as the standardization of different security sectors with respect to 5G privacy. This include network softwarization security and physical layer security. The security functionality of 3GPP 5G networks is discuss in \cite{x015} detailing security requirement, existing security solutions techniques in IoT devices, V2X, and D2D communication with new technical features in 3GPP 5G networks. Achieving low latency communication in three different domains (i.e., RAN, core network, and caching) is discuss in the reference \cite{011},\cite{001} and 5G cellular technology capable of meeting the latency requirement, such as SDN, NFV, caching, and mobile edge computing. H. Md Frahad, et al. \cite{xx015} surveyed the advance research on C-RAN for 5G cellular system focusing on various enhancement, and analysis to support the 5G heterogeneous cellular network taking into consideration related research works on throughput enhancement, latency, interference management, security and system cost reduction, as well as energy efficiency. Open RAN evolution from cloud-RAN is presented in \cite{xx016}, with a generic concepts and limitations in the evolution of C-RAN. The standardization efforts of V-RAN and Open RAN with potential development and research activities that are imminent to the C-RAN and its derivatives (i.e., V-RAN and O-RAN).
The role of open-source in 6G wireless networks is discuses in \cite{xx017}, detailing the challenges faced by Telecom Service Providers (TSPs) for innovative solutions in reducing the deployment cost. Open-source projects solutions such as OpenDayLight, OpenStack, M-CORD, ONARP and OPNFV are expected to play a vital role as well as dissaggregation of the RAN with AI integration. These collective initiative will drive the trend toward optimized solution in terms of cost and innovation which also rely on the SBA architectural concept based programmable plug and play open architecture for 6G standardization. The integration of AI/ML technique in O-RAN architecture is discussed in \cite{xx018} which can serve for optimization purposes and the use of white-box and open-source software enable new players to take part in the market share. The use of AI/ML in O-RAN allows business opportunities through the multi-vendor environment with improved system design. 
\subsection{Our Contribution}
In this survey, we broadly discuss how open-source software, open-source framework, and open-source project can be fitted into the CN, RAN, and edge applications as well as future evolution. We provide a comprehensive state-of-the-art literature review and taxonomy on security threats, such as control and data plane attack, IoT security attacks, CN attack, SDN/NFV DoS/DDoS attack, and RISC-V processor attack just to mention a few. Our contribution to the open-source-defined wireless networks in this paper is as follows:

1) Technicality of open-source-defined wireless networks security challenges and how to over come the problems, we further detail discussions on open-source software, open-source hardware, open-source project, and open-source framework, which are fitted into the CN, RAN, and edge network as well as implementation.

2) We provide comprehensive analysis of open-source-defined wireless networks trend in the coming future evolution to our readers and discuss the industrial security view on 5G, unfold several research capabilities, research efforts from both academia and industry on the impact of open-source-defined wireless networks.

3) Finally, we provide a conceptual federated SDN controller based open-source-defined wireless networks in Fig. \ref{fig11}, in which we stitched the 3 controller domain via west/east bound interface to create federated open-source-defined wireless networks. We also present lessons learned, open research issues and future research directions in adopting open-source-defined wireless networks into the communication domain. 

\subsection{Organization of This Paper} 

In this subsection, we discuss the organization of this paper which explore and promote the progress of open-source-defined wireless networks, via the key technology enablers and comment on their architectural software and hardware from the perspective of open-source-defined wireless networks. 
In this survey, we present the framework, key technology of open-source-defined wireless networks in Section \ref{BB} . We also discuss open-source-defined CN and RAN the former is based on SBA and the latter is based on disaggregation in Section \ref{CC} and \ref{AA} respectively, and open-source-defined MEC in Section \ref{EE} . In Section \ref{FF} , we present a comprehensive review focusing on challenges, security, current and future evolution regarding ongoing research efforts in open-source-defined wireless networks. We discuss open-source hardware and its classification in Section \ref{GG} . Section \ref{HH} present and demonstrate the implementation of this new paradigm. Finally, we conclude this paper and address the future potential work in Section \ref{II} . A summary of important acronyms that will be frequently used is provided in table \ref{tab.II} . We present our survey outline structure in Fig. \ref{fig1}.
\begin{table*}[]
	\caption{Summary of important acronyms }
	\label{tab.II}
	\centering
	\begin{tabular}{|c|c|c|c|}
	\hline
	\textbf{Acronym} & \textbf{Meaning}                                    & \textbf{Acronym} & \textbf{Meaning}                               \\ \hline
	ABNO             & Application Base Network Operation Architecture     & AP             & Application Protocol \\ \hline
	AMF              & Access Management Function                          & API              & Application Programmable Interface             \\ \hline
	ASF              & Application Selection Function                      & AI               & Artificial Intelligence                        \\ \hline
	CAPEX            & Capital Expenditure                                 & CeNB             & Control Plane eNB                              \\ \hline
	CO               & Central Office                                      & CORD             & Central Office Re-architect as a Data Center   \\ \hline
	CU               & Central Unit                                        & C-RAN            & Cloud-RAN                                      \\ \hline
	CLI              & Command Line Interface                              & DANOS            & Disaggregated Network Operating System         \\ \hline
	DU               & Distribution Unit                                   & EON              & Elastic Optical Network                        \\ \hline
	eNB              & Evolved NodeB                                       & EPC              & Evolve Packet Core                             \\ \hline
	xRAN             & Extensible RAN                                      & 5G               & Fifth Generation                               \\ \hline
	4G               & Fourth Generation                                   & gNB              & Next Generation NodeB                          \\ \hline
	GSM              & Global System for Mobile                            & HDD              & Hard Disk Drive                                \\ \hline
	HTTP             & hypertext transfer protocol                         & IM               & Infrastructure management                      \\ \hline
	IETF             & Internet Engineering Task Force                     & IoT              & Internet of Things                             \\ \hline
	L2               & Layer 2                                             & L3               & Layer 3                                        \\ \hline
	LTE              & Long Term Evolution                                 & M2M              & machine to machine                             \\ \hline
	MANO             & Management and Orchestration                        & MAC              & Media Access Control                           \\ \hline
	MCORD            & Mobile Central Office Re-architect as a Data Center & MEC              & Mobile Edge computing                          \\ \hline
	NFV              & Network Function Virtualization                     & NFVI             & Network Function Virtualization Infrastructure \\ \hline
	NFs              & Network Functions                                   & NIC              & Network Interface Card                         \\ \hline
	NRF              & Network Repository Function                         & NGF              & Next Generation Fronthaul Interface            \\ \hline
	OAM            & Operation Administration and Maintenance  & OCU   & Open Centralize Unit             \\ \hline
	OCP              & Open Compute Project                                & ODTN              & Open Disaggregated Transport Network                           \\ \hline
	ODL              & Open Day Light                                      & ODU              & Open Distributed Unit                          \\ \hline
	ONAP             & Open Network Automation Platform                    & ONIE             & Open Network Install Environment               \\ \hline
	ONL              & Open Network Linux                                  & ONOS             & Open Network Operating System                  \\ \hline
	O-RAN            & Open Radio Access Network                           & OSS              & open-source Software                           \\ \hline
	OAI              & OpenAirInterface                                    & OpenBSC          & Open Base Station Controller                   \\ \hline
	OpenBTS          & Open Base Transceiver Station                       & OPEX             & Operational Expenditure                        \\ \hline
	OSI              & Open-Source Initiative                         & OTT & Over The Top \\ \hline     
	PDCP             & Packet Data Convergence Protocol               & PGW-C     & Packet Gateway Control Plane\\ \hline
	PGW-U            & Packet Gateway User Plane                      & POC & Proof-of-Concept     \\ \hline
	QoS              & Quality of Service                             & RF  & Radio Frequency       \\ \hline
	R\&D             & Research and Development     & RLC              & Radio Link Control                             \\ \hline
	RRU              & Radio Remote Unit                                  & RAN              & Radio Access Network                           \\ \hline
	REST             & Representational State Transfer                     & SON              & Self-Organizing Network                        \\ \hline
	SBI             & Service Base Interface                              & SRF              & Service Registry Function                      \\ \hline
	SGW-C            & Serving Gateway Control Plane                       & SGW-U            & Serving Gateway User Plane                     \\ \hline
	6G               & Six Generation                                      & SDI              & Software Define Infrastructure                 \\ \hline
	SDN              & software define network                             & SDS              & Software Define Storage                        \\ \hline
	SD-WAN           & Software Define Wide Area Network                   & SDR              & Software Defined Radio                         \\ \hline
	SSD              & Solid State Drive                                   & SVM              & Support Vector Machine                         \\ \hline
	TIP              & Telecom Infrastructure Project                      & 3G               & Third Generation                               \\ \hline
	3GPP             & Third Generation Partnership Project                & 3D               & Three-Dimensional Communication                \\ \hline
	TLV              & Type-Length-Value                                   & UE               & User   Equipment                               \\ \hline
	UeNB             & User Plane eNB                                      & UPF              & User Plane Function                            \\ \hline
	vEPC             & Virtual Evolve Packet Core                          & VIM              & Virtual Infrastructure Manager                 \\ \hline
	VNF              & Virtual Network Function                            & vRAN             & virtual RAN                                    \\ \hline
	VoLTE            & Voice over LTE                                      & YANG             & Yet Another Next Generation                    \\ \hline YAML                           & Yet Another Markup Language & & \\ \hline 
\end{tabular}
\end{table*}
\section{Frameworks}\label{BB}
In this section, we discuss the frameworks of open-source-defined wireless networks. We further classify its categories into a black box, grey box, and white box based on the level of disaggregation and openness with either closed or open APIs. The first two are well studied with lots of arguments on the latter, therefore, we focus on the white box.
This survey reviews all the current related frameworks so far, including our demonstrated proof-of-concept framework as shown in Fig.\ref{fig2} depicting of user plane, control plane, and MANO plane, through the disaggregation of Network Functions (NFs). This provides the ability and capability of low latency, high bandwidth, and real-time access to the control plane resources. These disaggregated NFs life cycles can be managed and orchestrated through the MANO plane. Open-source-defined wireless network framework scheme improves openness and flexibility of traditional and proprietary systems of closed and vendor-legacy equipment.

\subsection{Open-Source-Defined Wireless Networking Frameworks}

In this subsection, we discuss the proof-of-concept (POC) frameworks which are based on key enabling technology of SDN/NFV and open-source Software (OSS) running on common hardware platform processors (e.g., x86 and ARM architectures). Such OSS infrastructure provides openness and modification to RAN, which cannot be possible with proprietary infrastructure. This is because proprietary cellular networks are not easy to deal with in developing and deploying edge applications because of their legacy interface. To solve this problem, OSS-base RAN upon the commercial spectrum with the ability of Development Operation (DevOps) can be used and applied. Such OSS can allow integration, upgrading, and modification of the system elements, such as Evolve Packet Core (EPC), and applications executed at the edge. The disaggregated Network Function (NFs) life cycle can be managed and orchestrated at the MANO plane. In the context of management and orchestration, a lot of open-source projects have put forward the proposed implementation of the NFV-MANO framework at a different level. This includes open-source MANO (OSM), Open Network Automation Platform (ONAP), and so on. Most of these projects are using OpenStack and recently Kubernetes (K8s) as their VIM. 
\subsubsection{Proof-of-Concept Architectural Framework}
This framework is made up of three building blocks, and these are VIM, Virtual Network Function (VNF) manager, and NFV orchestration. In this, OSM manages the scaling, creation, and deletion of VNFs by commanding the VIM on what to do \cite{42}, and put the northbound interface for different clients with future expectations. In addition, ONAP can also be a good candidate in designing, creating, orchestrating, and managing the life cycle of VNFs, as well as SDN \cite{43}. Mobile network operators can inject this into their network infrastructure to provide automation and orchestration to both application terminal, and infrastructure. The user plane function serves as a gateway to the public internet. 

Furthermore, as shown in Fig.\ref{fig2} the data traffic forwarding user plane is supported through the following layers:
\begin{itemize}
\item The infrastructure layer consists of computing resources such as the CPU, caching resource, e.g., memory and hard disk, and a combination of communicating resources, e.g., Network Interface Card (NIC) and bandwidth. These resources can be located remotely at a geographical location which can be distributed across many locations.
\item Virtualization layer:
The virtualization layer decouples the tightly couple resources of the infrastructure layer (i.e., storage, compute, and communication) from their dedicated hardware into VNF, which is accomplished by the aid of NFV technology. In this layer, the hypervisor allocates decouple NFs to all Virtual Machines (VM). This enables each VM to get its share from the physical decouple resources (i.e., CPU, memory, NIC, etc.) which is a traditional method of virtualization. Modern virtualization now consists of a docker engine that eliminates the use of hypervisor and VM. Moreover, Docker containers are used instead of VM. The former (i.e., docker container) are very light, fast to boot and execute because containers run the operating system like the host computer, whereas VM runs a different operating system \cite{45}. Furthermore, our framework containers provide an isolated environment to prevent interference with edge applications, as well as multiple operating systems by the host computer at the edge. Therefore, making the infrastructure more flexible and maintainable. The framework allows containers to be managed through management and orchestration software that automates the creation, deletion, upgrading, restoring, and so on. 
\item Control layer: 
The control layer provides network control to services via a service-based layer for decoupling NFs to communicate to one another with SBI protocol conforming to the web protocol, i.e., HyperText Transfer Protocol (HTTP). Latency minimization, offloading traffic, and service deployment can be among the services users can benefit from the design framework. In addition, traffic offloading brought to the network edge helps in optimizing and minimizing the CN's loads with minimum latency and delay, when utilizing the services at the edge \cite{46}.
\begin{figure}[t]
	\begin{center}
		\includegraphics[width=3.5in]{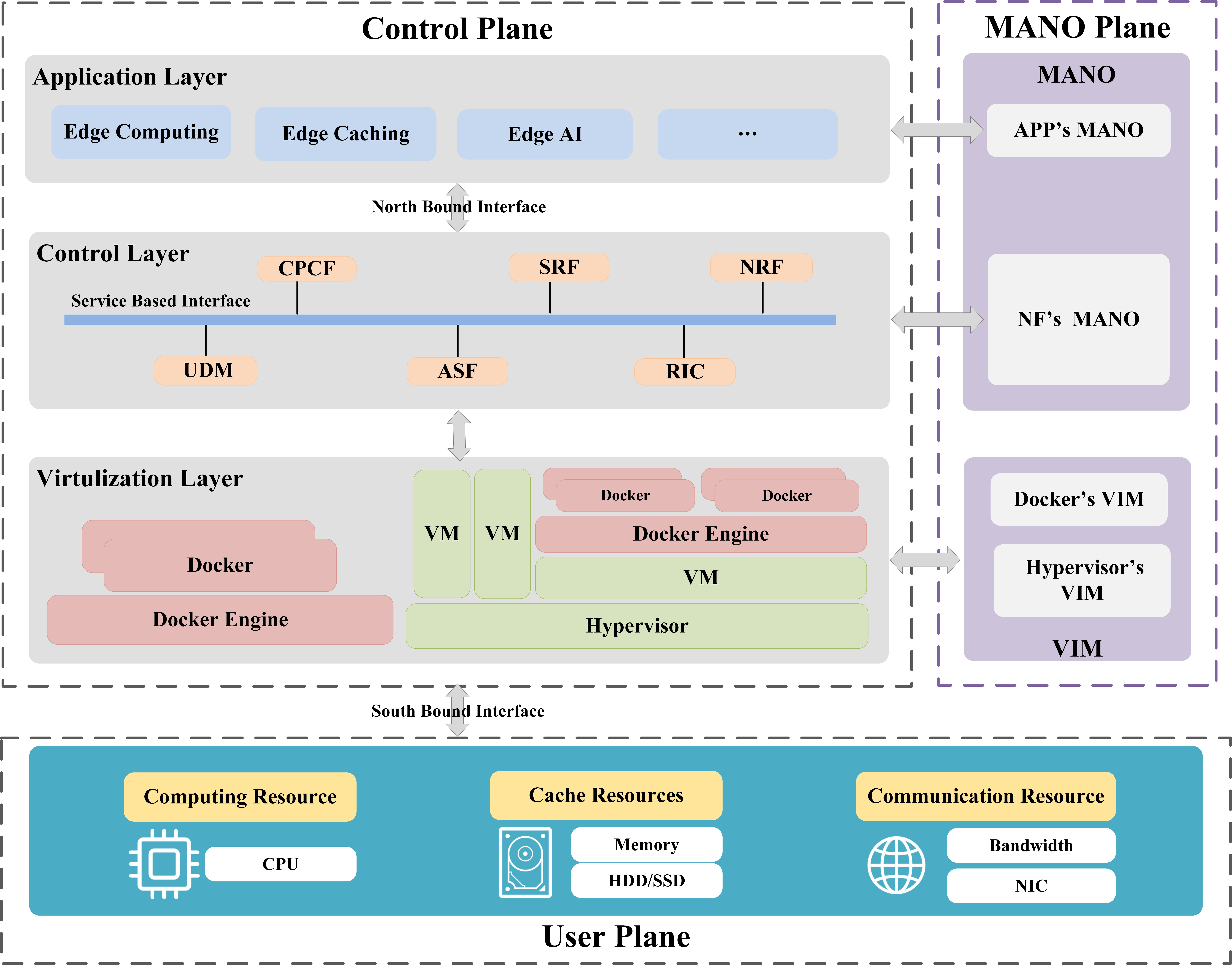}
		\caption{Proof-of-Concept framework}
		\label{fig2}
	\end{center}
\end{figure}
\item Application layer:
Application development at the edge is difficult due to the proprietary interfaces, as such a container-based application can be used in developing the edge application and executed without difficulty. Based on our framework, the application layer consists of resource allocation, AI-enable traffic identification, and network slicing. Container applications such as docker can provide easy development, deployment, and execution of the distributed application. Moreover, the container does not require a full launch of the operating system for each application to be executed. Developers can take advantage of containers to overcome all the vendor-specific applications that are developed from different operating systems and different platforms \cite{47}. On the other hand, the MANO plane orchestrates the VNFs and the VIM that manages the virtualized resources.

\end{itemize}
\begin{table*}[]
	\caption{Summary of open-source project and their use cases }
	\label{tab.III}
	\centering
		\begin{tabular}{|c|c|c|c|c|c|}
			\hline
			\textbf{Project} &
			\textbf{Focus} &
			\textbf{Organization} &
			\textbf{VIM use} &
			\textbf{\begin{tabular}[c]{@{}c@{}}Data modelling \\ language use\end{tabular}} &
			\textbf{Use case scenario} \\ \hline
			ONAP &
			\begin{tabular}[c]{@{}c@{}}Network management\\  and Orchestration\end{tabular} &
			Linux foundation &
			OpenStack &
			TOSCA yaml &
			5G and VoLTE \\ \hline
			OSM &
			\begin{tabular}[c]{@{}c@{}}Network management\\ and orchestration\end{tabular} &
			ETSI &
			\begin{tabular}[c]{@{}c@{}}OpenStack,\\ Vmware,\\ OpenVIM\end{tabular} &
			YANG yaml &
			\begin{tabular}[c]{@{}c@{}}5G slicing, \\ service chaining, \\ and MEC\end{tabular} \\ \hline
			OPNFV &
			\begin{tabular}[c]{@{}c@{}}Infrastructure\\ management\end{tabular} &
			Linux foundation &
			\begin{tabular}[c]{@{}c@{}}Kubernetes and\\  OpenStack\end{tabular} &
			YANG yaml &
			vEPC \\ \hline
			Cloudify &
			\begin{tabular}[c]{@{}c@{}}Network management\\ and Orchestration\end{tabular} &
			\begin{tabular}[c]{@{}c@{}}Cloudify \\ platform Ltd\end{tabular} &
			\begin{tabular}[c]{@{}c@{}}OpenStack,\\ Vmware, \\ and Docker\end{tabular} &
			YANG yaml &
			SD-WAN and vEPC \\ \hline
			M-CORD &
			\begin{tabular}[c]{@{}c@{}}Service chain and\\ real-time resource \\ management\end{tabular} &
			Open networking lab &
			OpenStack &
			P4 &
			\begin{tabular}[c]{@{}c@{}}Edge cloud mobile \\ network\end{tabular} \\ \hline
            O-RAN &
            \begin{tabular}[c]{@{}c@{}}5G radio access network\\ implementation\end{tabular} &
            O-RAN alliance &
           \begin{tabular}[c]{@{}c@{}} OpenStack, Vmware, Docker,\\ container\end{tabular} &
            YANG yaml &
            \begin{tabular}[c]{@{}c@{}}RAN slicing\end{tabular} \\ \hline
		\end{tabular}
	\end{table*}

\subsubsection{O-RAN Framework}
\begin{figure*}[t]
	\begin{center}
		\includegraphics[width=5.5in]{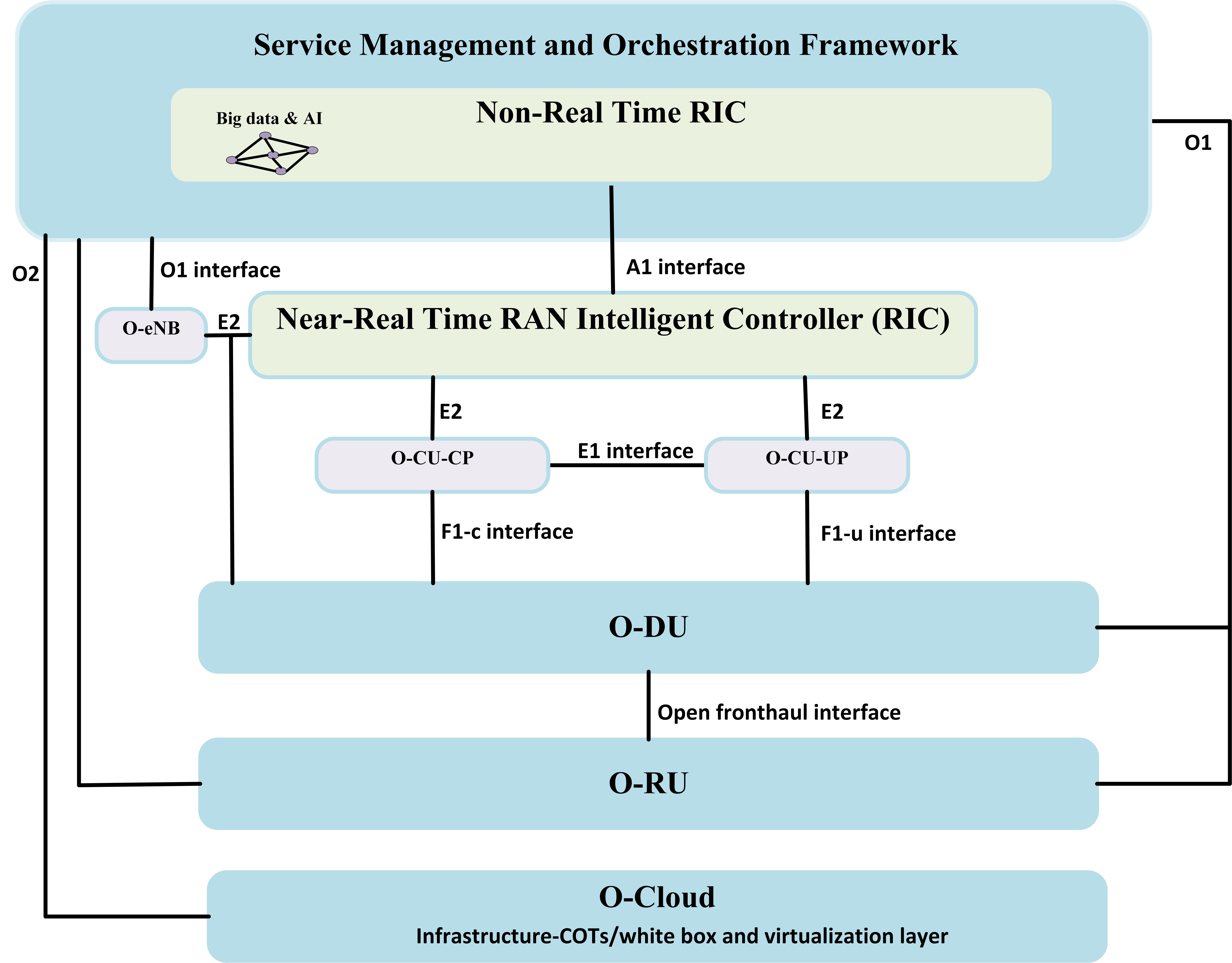}
		\caption{O-RAN architectural logical view}
		\label{fig3}
	\end{center}
\end{figure*}

As shown in Fig.\ref{fig3}, the O-RAN framework is defined on disaggregation of RAN with the introduction of new open interfaces from the group alliance (i.e., AT\&T, china mobile, Deutsche telecom, NTT Docomo, and orange), to promote flexibility, speed, and innovation to 5G RAN. This has led both the O-RAN alliance and the linux foundation to focus on software implementation that is compatible with the open architectural specifications to facilitate open implementation. The main goal is to achieve a promising solution that can be used in unifying and accelerating the deployment and evolution of the RAN \cite{x001}. Intelligence in the RAN is achieved by leveraging deep learning techniques from AI to be embedded in the RAN layer architecture. 
Big data and AI can pave the way for the telecom industry to optimize their network, better user experience, capacity improvement, and give room for more innovation, with decades of conventional telecom wired experience, now shifting to the wireless paradigm. With the recent progress of the introduction of CU and DU in the 5G networks as well as edge computing, this gives room for Big Data Analytic (BDA) to be introduced into the RAN.

However, Machine Learning (ML) technique can be used in problem-solving in wireless communication as well as in MAC layer \cite{127}. The BDA enhances resource orchestration, RAN optimization, and content distribution. With seamless database management of data collected, wireless BDA will be able to process the data and classify it according to the fundamental features of each particular data, incorporating with ML technique, such as Support Vector Machine (SVM), Self-Organizing Network (SON), and full self-driven network. Prediction and prescription can be performed, unlike conventional networks in which collection and processing of the data are difficult and a burden to the network \cite{128}. The BDA if applied to traffic, signaling, as well as environmental data, can assist autonomous network optimization in establishing a big database network operation. 
This can contribute in analyzing the traffic which provide the fundamental for dynamic network deployment \cite{129}. 

Zahid et al. \cite{016} focuses on the state-of-the-art application of BDA in telecommunication with the proposal of LambdaTel architecture that is completely open-source-based BDA technologies (for more information on LamdaTel see \cite{017}), from the perspective of BDA architecture and infrastructure of the telecom industry. This stems from the recent advancement in the BDA to realize novel opportunities accredited from telecom big data, that motivate the authors to investigate the fast-changing in BDA technology landscape applicable to the telecom sector. The BDA telecom application is systematically reviewed with more than thirty articles group according to the framework, literature review, use cases, white papers, and experimental validation. However, the experimental validation all present PoC on the BDA technology stack with no full BDA implementation in the telecom ecosystem.

Furthermore, the work in \cite{x002} implement AI/ML workflow, defined by work group 2 (WG2) AI/ML specifications with open-source software (i.e., ONAP and Acumos) from the software community of O-RAN. 
Based on the framework of open-source software acumos, the authors generate and package ML models to be executed in RAN Intelligent Controller (RIC), with components of ONAP providing monitoring capability and arbitration required to implement the workflow. This is to facilitate the close monitoring of the RAN resources, automation, and optimizing the RAN with RIC through software and ML techniques, such as reinforcement learning, federated learning, and Graphic Processing Unit (GPU) which are hardware accelerators. The success of ML enables AI with complex problems and tasks been solved, such as image recognition ML has attracted the attention from both academia and industry in the implementation of ML in wireless networks.

Sun et al. \cite{x008} surveyed the recent progress and application of ML in wireless communication according to MAC layer, network layer, and application layer with each layer hosting a particular task, such as resource management, network, and mobility management, and localization respectively. However, ML-based network management and mobility defined the base station switching control, applications clustering, user association, and routing, with identified conditions on when to apply ML in wireless networks and the choice of ML technique to be adopted by the user in comparison to tradition method applied to wireless network communication. In fact, infrastructure updates and open data set are needed in overcoming the challenge of implementing ML in wireless communication.
   
The integration of O-RAN with MEC, network slicing, and Self-Support-Network (SSN) to solve issues that are not properly addressed by O-RAN is presented in \cite{x003}. Moreover, these technologies are complementary and overlapping. The integration provides synergy between the technologies, such that a developing mechanism in one technology can serve as a solution to another. In an effort to demonstrate the relaxation of fronthaul in 5G network with O-RAN, Gomes et. al \cite{x004} present the amount of Resource Block (RBs) required per user to deploy the minimum amount of data rate (i.e., 100 Mb/s) per user based on the functional split. The computed required minimum value of RBs is then compared with the fronthaul traffic demand of O-RAN with respect to the traffic of conventional CPRI which proves to optimize 
the traffic compared to the legacy CPRI fronthaul. Furthermore, O-RAN disaggregation, self-driven, virtualize, and SDN has made it the best candidate in support of IoT devices and high-speed application that requires low latency \cite{x005}. The O-RAN framework has attained the success of decoupling RAN hardware from the functional software to implement auto-scaling and reconfiguration of RAN NFs, from the surging demand of data traffic. With current network densification to provide coverage and increase capacity, this results in cells number increase and therefore become more difficult to manage and optimize cell neighbor relationship.

This motivated the work in \cite{x006} to investigate a novel approach with respect to Automatic Neighbor Relationship (ANR), which is a recognized SSN function in both management of cell neighbor relationship and optimizing the cell neighbor relationship table, that improves successful handover timing and minimizes call drop. The authors approach ANR optimization with O-RAN open interfaces (e.g., E2) and architecture to leverage non-real-time RIC in support of intelligent models, such as ML with proactive ANR optimization to enhance the gNodeB handover.

The NR success was achieved by splitting the higher layers of the 3GPP stack (PDCP, SDAP, and RRC) and the lower layers (RLC, MAC, and PHY) into two successive different logical units i.e., centralized unit (CU) and distributed unit (DU), as well as the radio unit (RU) from the physical layer lower part in a standalone deployment as discuss in Section \ref{AA} . The O-RAN implement interfaces which include the A1 \cite{x106} that connect the two RICs (i.e., non-real-time RIC and near-real-time RIC). The A1 interface enables and manages the ML models as discussed earlier. The O1 \cite{x107} interface enables the interaction with the O-RUs and permits the non-real-time RIC to perform operations which include communication management and fault supervision. The E2 \citen{x108} interface extend from the near-real-time RIC to O-CU, O-DU, and O-eNB and coordinates the functional control and management of the node's interaction with the RIC. The E1 interface extends from the control to the user plane, whereas the F1 interface implements the F1-control and user place which operate between the O-DU and O-CU respectively. Both E1 and F1 conform to the 3GPP specifications. Finally, the O-cloud provides the fundamental physical infrastructure nodes to realize the O-RAN hosting requirement of functions, such as near-real-time RIC, O-CU-CP, O-CU-UP, and O-DU including the software components (i.e., VMs, operating system, and so on) \cite{x109}.    
\subsubsection{Telecom Infra Project}
Telecom Infra Project (TIP) is a global community collaboration founded in 2016 with a vision to transform the closed and proprietary telecom infrastructure into the open, interoperable, and multi-vendor product to accelerate innovation, deployment, and time-to-market. This project envisioned addressing strategic network areas, such as access network, transport network, and core network\cite{a006}.
\begin{itemize}
	\item RAN project group: The RAN TIP project focus on the improvement of RAN infrastructure's efficiency and flexibility enable through software-based designed and open architecture for flexible and agile RAN equipment, which includes OpenRAN, vRAN fronthaul, OpenCellular, OpenRAN 5G NR Base station, and CrowdCell.
	\item Transport project group: This project group’s goal is to achieve fast, scalable, and ease of configuration in open optical and packet transport, and over the extensibility challenge in wireless backhaul. Open optical and packet transport engineering groups focus on implementing network open architecture, such as optical transponder, open line system, and open interfaces in optical and IP networks. 
	\item Open CN project: The open CN project group is determined to implement an open and cloud-native core network that run on standardized hardware and software, with the collection of microservices which implement various network functions that are flexible and extensible.
\end{itemize}
\subsubsection{Open Programmable Secure-5G (OPS-5G)}
Defined by Defense Advanced Research Project Agency (DARPA) to deliver cost-effective and secure 5G/6G mobile networks with open-source software and hardware, for fast and easy deployment with transparency from open-source technology which serves as a primary element of OPS-5G. This provides the ability to examine and modify the source code for the desired purpose with easier porting and portability from devices, to decouple the software and hardware. Addressing 5G security is a priority from IoT devices to servers, such as side-channel attacks introduce from virtualization and network slicing through programmability to enable security\cite{b006}. 

The OPS-5G envision to address this through four technical areas that include: (1) The acceleration of open-source software developed with online 5G machine translation of standard documents. Through the cooperate existence of Independent Test and Evaluators (ITE) with technicality; (2) Implementing security technique and security architecture, to enable security across the entire devices; (3) Network slice isolation to prevent security risks from untrusted infrastructure, because network slice overlays virtual networks across different enterprises. The use of programmable networks to shape the selection of physical components for slice isolation approaches at both system and network-level can be accomplished by route separation to minimize the use of untrusted resources; (4) The principal idea and technique of programmable network by OPS-5G increase security and innovation via the programmable element of OPS-5G infrastructure against compromising element, not in OPS-5G software control.

This is demonstrated in\cite{c006} which open and programmable 5G network-in-box is implemented to provide 5G private network deployment for industries, base open-source software stack, and general-purpose hardware which operate in 5G Non-StandAlone (NSA) integrating 4G long-term evolution (LTE). This showcases the integration of white-box 5G solutions based on open-source software and general-purpose hardware.
\subsubsection{Open Networking Foundation (ONF)} 
Open Networking Foundation is a non-profitable consortium of telecom operators contributing to the open-source community for use in their wireless network domain. This has led to the successively open-source project innovations, such as ONOS, CORD (e.g., M-CORD, E-CORD, R-CORD) for operators built, deployed, and used the solutions in their PoCs reference design. The ONF envision innovation through disaggregation and open-source SDN/NFV platforms. Furthermore, ONF aims interoperability of interfaces with software-defined standards to enable a multi-vendor components ecosystem of plug-and-play fashion\cite{d006}. In an effort from ONF to promote multi-vendor products to operator base SDN and disaggregation. ONF presented SEBA (i.e., SDN enabled broadband access) that is fully open-source and built upon white-box, merchant silicon ASIC, and SDN principle. Mobile operators leverage the benefits of virtualization and cloudification of Passive Optical Network (PON) based broadband access network deployment of FTTB (i.e., Fiber-to-the-Building) through white-box hardware with three control and manageable software project, VOLTHA (i.e., Virtual Optical Line Terminal Hardware Abstraction), Trellis, and SD-BNG (i.e., Software-Defined Broadband Network Gateway) \cite{030}.
\subsubsection{Open5G Community} 
We discuss Open5G from our perspective as a cellular network that serves diverse heterogeneous use cases, such as mobile broadband users, ultra-low latency, and massive machine-type communication scenarios. This diverse required communication type will depend on open networks, open API, open-source software, open-source hardware, virtualization, and disaggregation of software/hardware. In realization of Open5G, technology enablers which include SDN as the candidate technology to stir the data application traffic from the time-varying topology due to its openness, well-defined interfaces, and programmability integrated with NFV technology to support network virtualization. From both CN base service-based architecture, to open radio access network-based disaggregation components such as white-box Radio Remote Unite (RRU), open CU, \cite{i006} open DU and open API as detailed in Section \ref{AA}-B.
\subsubsection{Other Frameworks and Projects in Open-Source-Defined Wireless Networks}
\begin{itemize}
\item Magma: Magma is an OSS framework initiated by Facebook for the purpose of deployment of cellular networks in remote areas, and to avoid vendor lock-in to operators via dependence on unique access type technology, such as LTE and WiFi. Magma consists of two building block components, which include an access gateway interfacing the CN, a cloud-based orchestrator which secures and monitor the network\cite{e006}.
\item LL-MEC: Low latency Multi-access Edge Computing (LL-MEC) platform is open-source which enables mobile network motoring, programmable, and control with compliance to the 3GPP and ETSI specification. LL-MEC leverage SDN and edge computing, providing an end-to-end service platform to the edge users. The capability of LL-MEC support for network slicing is presented in \cite{f006}, offering different latency requirements to diverse applications. 
\item COMAC: Converged Multi-Access and Core (COMAC) is an open-source project which extends CORD, OMEC, SEBA, and VOLTHA (Virtual OLT Hardware Abstraction) that enables the abstraction of hardware for broadband access with support for the multi-vendor product via disaggregation. The main goal of COMAC is the convergence of broadband access, CN, 5G networks, SDN, and WiFi technology.  However, COMAC leverage the O-RAN controller as an edge for mobile cellular access\cite{g006}.
\item Central Office Re-architect-as-a-Data Center (CORD): The open-source ONOS controller experiment resolves the Central Office Re-architect as a Data Center (CORD) or Mobile Central Office Re-architect as a Data Center (M-CORD), which both exists as a used case of ONOS. Mobile operators can provide end services to consumers through CODE. In addition, CODE provides significant benefits, such as Internet-as-a-Service (IaaS), Subscriber-as-a-Service (SaaS), and Monitoring-as-a-Service (MaaS). CORD is significant in such a way that it provides low cost compared to traditional operator's data center, through OSS and white box. However, Mobile-CORD (M-CORD) can provide a solution in the upcoming future data traffic. CORD \cite{120} is a transformable effort of changing the legacy Central Office (CO) in telecommunication network close and proprietary hardware in re-architect CO, replace with software that is executed on switches and commodity servers. This software is managed, customize, and orchestrated to benefit CO from both CAPEX and OPEX. In fact, CORD implementation has also focused on Residential CORD (R-CORD), M-CORD, and Enterprise-CORD (E-CORD). In CORD structural framework, XOS takes care of assembling the multi-tenant and is a CORD controller. ONOS provides the hosting of the application, OpenStack/Docker coordinates, and manages the compute instances. Moreover, the speed at which the mobile operator's data center is growing post a trend toward which the need of an operating system that is more scalable, low in cost, able to cope with the diversity of workload, performance, and efficiency \cite{122}.
Modern mobile operators' data center servers required an operating system that is application-defined, and this is XOS. XOS leverage the hardware in support for virtualization and taking out the resource management function from the traditional kernel to particular user space and provide application performance on bare metal. The requirement of low latency has resulted in decentralizing cloud computing into edge and fog computing with the idea of CORD. 
Yuang et al. \cite{032} presented Edged Data Center (EDC) network architecture with prototype testbed implementation, called Intelligent-defined OPtical Tunnel Network System (OPTUNS). OPTUNS enable the collection of packet transport through wavelength-based optical tunnel and consist of optical switching subsystems that facilitate packet control. However, with the targeted goal of achieving high bandwidth and ultra-low latency, OPTUNS reuses the massive wavelength with proactive optical tunnel control and fault tolerance. These optical tunnels are then controlled and coordinated through an SDN-based intelligent tunnel control system, with OPTUNS testbed prototype optical switching subsystem can interconnect hundreds of servers. In addition, power saving is achieved with OPTUNS to 82.6\%, contrary to the electrical spine-leaf networks in the data center.

\end{itemize}

\subsection{Categories of Open-Source-Defined Wireless Networking Frameworks}
In this subsection, we discuss the categories based on their level of openness into black-box, gray-box, and white-box networks. Whereby more emphasis is given on white-box network which is used in the carrier-grade network, and the profound impact it has on many segments and the open capability it enables through the network, and resource sharing can be identified.
\subsubsection{Black-box Network with Closed Capability} 
The close network capability of black-box can be described from four perspectives: (1) Black-box network implements a closed API with little or no innovation and ties an operator to a single vendor legacy product management and operation, with third-party application restriction. However, SDN technology enables open interfaces with selective APIs at both data and control plane;
(2) Closed network operating system (NOS) capability forces operator to choose and work with a close NOS, which internal details is not visible and impossible to modify according to the operator's choice and demand. This proprietary NOS is couple to it hardware which increase OPEX of service providers, with little or no freedom when new service upgrade required by the operator but have to spend weeks or months from product vendor; (3) Closed application-specific integrated circuit (ASIC) capability implement a proprietary ASIC coupled with both NOS and hardware from a single manufacturer. Each ASIC is designed to be custom for a specific purpose (e.g., some ASIC are best design for data centers). The close ASIC capability prevent resource and network optimization by a mobile operator in combining the designed custom specific ASIC suite needed on-demand service; (4) Closed software testing capability:
For every software to function properly testing phase is important. Software testing capability in black-box detail the working of the software, without access to the source code. Such a philosophy of close proprietary vendor solution stifle innovation and prevent new players into the game.

\subsubsection{Half-open and Half-closed Gray-box Network Capability}
The gray-box offers partial decisions and choices. This allows the mobile operator (e.g., China mobile, AT\&T, Airtel, etc.) to introduce the third-party application from custom design manufacture's NOS specification (e.g., Huawei, Samsung, Nokia, etc.), and contribute to the open-source community project suiting their demand. However, grey-box software testing capability provides the knowledge and details of the source code without modification.  
\subsubsection{White-box Network Capability}
\begin{figure}[h]
	\begin{center}
		\includegraphics[width=3in]{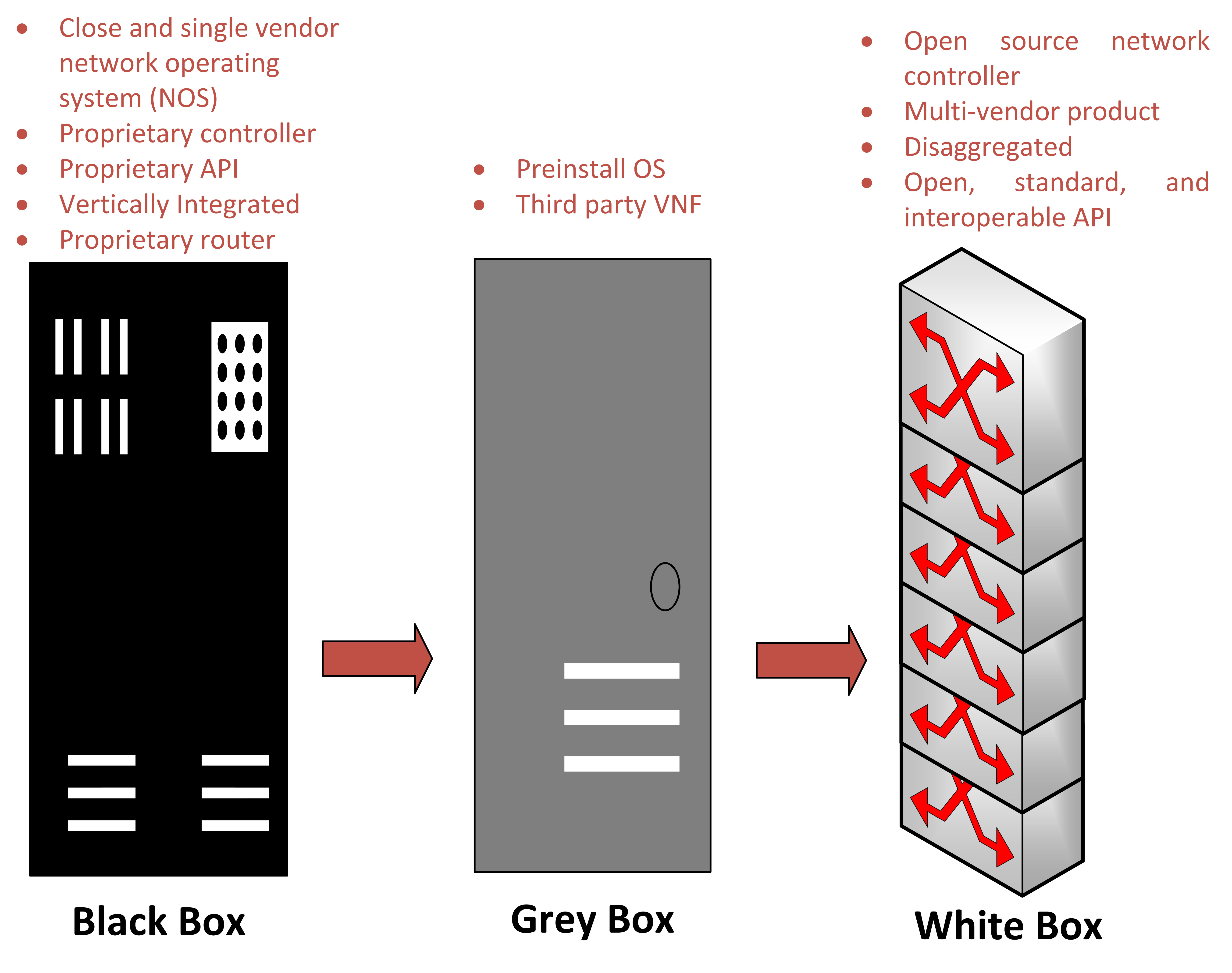}
		\caption{Transition state from black-box to white-box network}
		\label{fig5}
	\end{center}
\end{figure}

\begin{figure}[h]
	\begin{center}
		\includegraphics[width=3.5in]{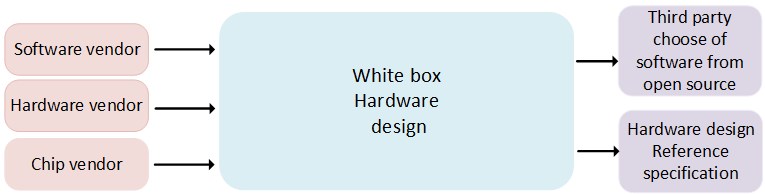}
		\caption{White-box network design}
		\label{fig6}
	\end{center}
\end{figure}

The goal of achieving a multi-vendor supply chain to prevent vendor lock-in and new players to commit in the market share, white-box open capability can be explained in three perspectives; (1) open-source software capability support for open-source software enables operators to develop independent source code that suite their need, based on open-source community code which required R\&D with technical efforts and investment. For example, Over-The-Top (OTT) vendors developed fully independent private cloud businesses with OpenStack. Fig.\ref{fig5} shows the transition trend from black-box to white-box network.
Open Network Linux (ONL) \cite{51} is an effort developed by OCP, SONiC \cite{52} and FBOSS \cite{53} are developed by Microsoft and Facebook respectively, and used in their data centers.  A DevOps concept is implemented giving the operators full control of their network traffic, and the software that runs on the need to be flexible, intelligent, self-driven, and interoperability; (2) Open hardware capability provide customize generic-off-the-shelf required bare-metal often used in SDN data plane which provides lower CAPEX and OPEX through disaggregation of hardware and software. 
This will enable service providers to deployed infrastructure at low cost with vendor lock-in prevention \cite{039}, and design modules from a variety of vendors, e.g., chip vendor, hardware vendor, and software vendor. Fig. \ref{fig6} depicts a white-box switch with a different component from a variety of vendors.


As a key factor to consider, the study shows that conventional data centers with traditional-based servers are cost to maintain considering a number of parameters. For example, life-cycle management cost makes the difference between the traditional data center and a data center-based white-box designed network, saving up to 40\% CAPEX/OPEX\cite{65}. 
Management and control of a white-box network can be archived through Network Configuration Protocol (NETCONF) and Yet Another Next Generation (YANG) model with support to SDN and automation provisioning. YANG model is a data modeling language that describes the data sent across the network management protocol, such as NETCONF which operators can use to configure and model the state of data across their wireless network.

\begin{figure}[h]
	\begin{center}
		\includegraphics[width=3in]{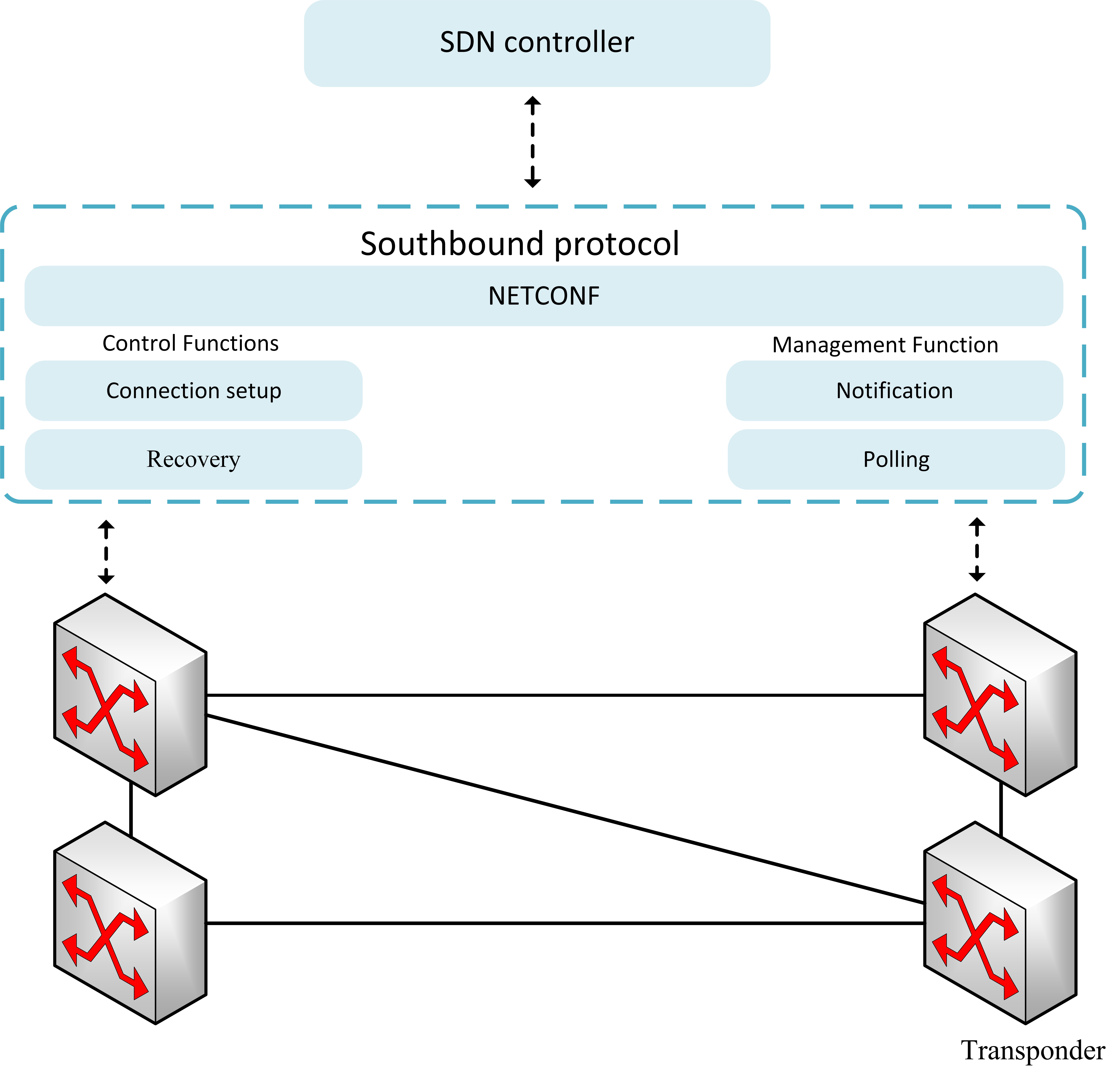}
		\caption{Demonstration of NETCONF in SDN environment}
		\label{fig7}
	\end{center}
\end{figure}

NETCONF is a protocol used in control and management of optical network in an SDN environment, enabling the configuration and access monitoring information of data plane devices \cite{68}. NETCONF leverage the YANG model to provide control and management of network elements that are independent of vendor. The experimental demonstration has taken place from both academia and research centers, in proving the capability of NETCONF and YANG model to analyze optical transponder support for variable rate and monitoring functionality \cite{69}. This led to optimum usage of optical spectrum base optical requirement, with bandwidth problem resolved and development of Elastic Optical Network (EON) is archived. 
Transponders are the core features of EON with a variety of functions that include determining the variation of bit rates and varying modulation schemes, that allowed mobile operators to optimize their network capacity. Data plane transponders can be configured to provide the required transmission properties, such as bit rate. The Internet Engineering Task Force (IETF) represent an open international community of network designers, vendors, operators, and researchers that work on the standard and technicality of the internet and are responsible for standardization of NETCONF as well as the proposal of Application Base Network Operation Architecture (ABNO). However, the management plane required standardization of different vendor devices in Operation Administration and Maintenance (OAM), with ABNO designating the functional entity of OAM handler \cite{66}. Fig.\ref{fig7} shows the SDN controller, data plane connection as well as transponder setup.

The concept of computes disaggregation within the centralized softwarised RAN is proposed in \cite{028}, which permit the allocation of process functions to individual server based on the required processing capacity. Through experimental analysis of BBU processing requirement with LTE open-source for wireless networks and linear programming modeling framework, results obtained show important benefits in energy saving contrary to the conventional method used. 
 
\begin{table*}[]
	\caption{Summary of open-source networking operating system}
	\label{tab.IV}
	\centering
	\begin{tabular}{|c|c|c|}
		\hline
		\textbf{Type}         & \textbf{Brief description}                                                                                                                                       & \textbf{Focus}                                                                                                                                              \\ \hline
		Stratum \cite{54}               & \begin{tabular}[c]{@{}c@{}}Development of a reference \\ design for white box switches \\ that support next generation SDN\end{tabular}                          & \begin{tabular}[c]{@{}c@{}}To avoid vendor lock-in \\ of current data plane\end{tabular}                                                                    \\ \hline
		Cumulus \cite{55}              & \begin{tabular}[c]{@{}c@{}}A Linux Debian base that run \\ on variety of commodity hardware\end{tabular}                                                         & \begin{tabular}[c]{@{}c@{}}Contribution in to multiple project \\ including Open Network Install Environment \\ (ONIE) to open compute project\end{tabular} \\ \hline
		Big switch light  \cite{56}    & Develop to integrate with white box hardware                                                                                                                     & Cloud networking                                                                                                                                            \\ \hline
		Open network Linux \cite{57}    & \begin{tabular}[c]{@{}c@{}}A bare metal switch forwarding \\ design for commodity component\end{tabular}                                                         & Open networking hardware                                                                                                                                    \\ \hline
		Dent Linux foundation \cite{58} & \begin{tabular}[c]{@{}c@{}}Utilization of Linux base project for \\ standardization of networking operating system\end{tabular}                                  & \begin{tabular}[c]{@{}c@{}}Enabling of the transition into disaggregated \\ networking component\end{tabular}                                               \\ \hline
		PicOS   \cite{59}              & \begin{tabular}[c]{@{}c@{}}Design to run on number of bare metal \\ switch from different vendors\end{tabular}                                                   & \begin{tabular}[c]{@{}c@{}}Assist in Switching and forwarding \\ of existing network\end{tabular}                                                           \\ \hline
		FTOS   \cite{60}               & \begin{tabular}[c]{@{}c@{}}Develop by Force10 network and \\ the use with open hardware\end{tabular}                                                             & Networking hardware devices                                                                                                                                 \\ \hline
		OcNOS \cite{61}                 & Used for data center and enterprises                                                                                                                             & Disaggregated networking                                                                                                                                    \\ \hline
		Coriant       \cite{62}        & \begin{tabular}[c]{@{}c@{}}Coriant which is targeted at service \\ providers and carriers that are interested \\ at disaggregation\end{tabular}                 & \begin{tabular}[c]{@{}c@{}}SDN/NFV solution, data center interconnect, \\ open line system\end{tabular}                                                     \\ \hline
		OpenSwitch   \cite{63}         & \begin{tabular}[c]{@{}c@{}}A community base network operating system \\ for disaggregated switches built in compliance\\  with open compute project\end{tabular} & \begin{tabular}[c]{@{}c@{}}Focus on full properties of the control plane with \\ L2, L3 network protocol features\end{tabular}                              \\ \hline
	\end{tabular}
\end{table*}

\section{Open-Source-Defined Core Network}\label{CC}

In this section, we discuss the Service Base Architecture (SBA) whose goal is to provide software-defined and programmable CN, with control plane functionality based on interconnected NFs access and delivery of individual network services through a service-based interface. Furthermore, we discuss MultiFire-based and CN distributed management as defined by 3GPP.

\subsection{Service Based Architecture (SBA)}
In this subsection, we discuss the design technicality of SBA that aims to support the diverse network requirement from both operators and the vertical industry with flexibility, efficiency, and openness to the CN. Mobile operators leverage the advantage of SBA that lays the future cloud infrastructure to their cloud applications. This enables the telecom industry to shift away from communication protocol-centric to a more modularized form of service i.e., service-based with modules of NFs that interact with other NFs through SBI. The SBI interaction with NFs conform to the web service REST (REpresentational State Transfer) or more commonly referred to as Restful state, is a protocol used on the internet. The inelastic architecture of the current mobile CN which continues to increase CAPEX and OPEX of service providers, built on dedicated expensive hardware racks and located at a fixed geographical terminal, forces operators to find a way out from circle centric of single-vendor services and management \cite{70}. Open-source-defined CN provide the absolute solution that allows flexible deployment of virtual mobile network function on general-purpose hardware, with full control and routing policy managed in a centralized manner through the concept of SBA, that provide an atomized service capability in 5G network.

\subsubsection{Software-Defined}
\begin{figure}[h]
	\begin{center}
		\includegraphics[width=3in]{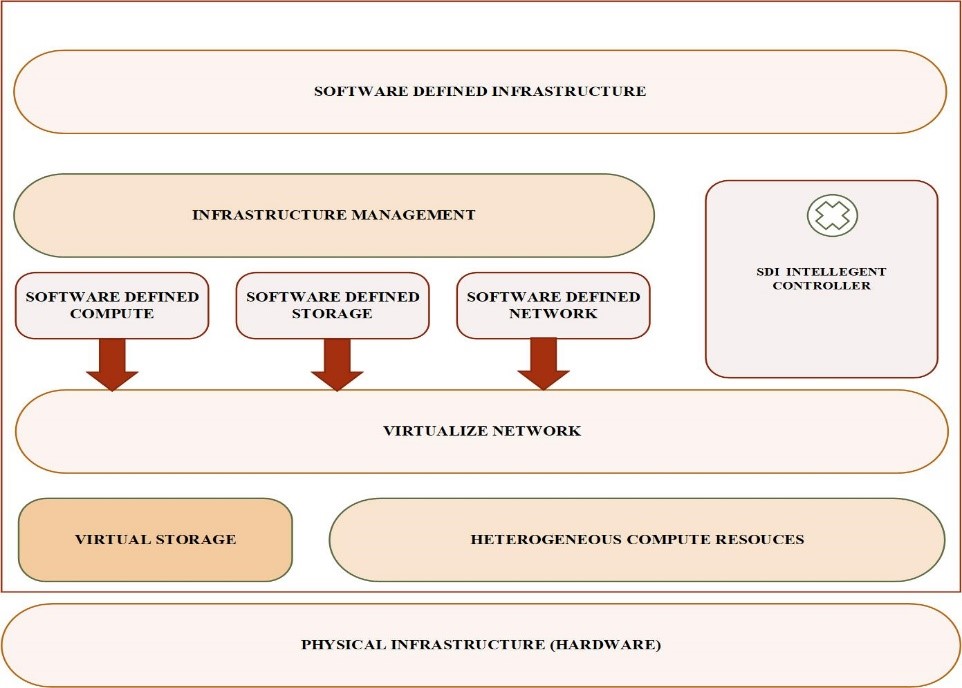}
		\caption{Software defined infrastructure with intelligent controller}
		\label{fig8}
	\end{center}
\end{figure}

The concept of virtualization has changed the perception of infrastructure with a great momentum change. As a result of the consumption of virtualized resources, a shift to the concept of Software-Defined Infrastructure (SDI) is needed which gives the consumer the ability to just a click away fashion. The SDI economic importance, such that there is no CAPEX, only OPEX\cite{72}. SDI is the infrastructure that keeps transforming itself by leveraging heterogeneous compute resources, in-contrast to conventional infrastructure, in SDI all elements are expected to be automated with each silo of (storage, compute, infrastructure) becoming a fundamental component of automation \cite{75}. However, the mapping of virtual infrastructure to physical infrastructure requires an intelligence controller just as depicted in Fig. \ref{fig8}. 

In addition, the Infrastructure Management (IM) can be a piece of software that provide an interface to the mobile operator. The require infrastructure properties, such as scalability, availability, and upgradability. Moreover, SDN decouple the control plane from the data plane, in same fashion software define storage (SDS) decouples the control and management plane from the data plane of a storage device. The software define environment integrate the control and management plane of software-define compute, storage, network and as well as infrastructure. The computing infrastructure (i.e., compute, storage, and network) is transforming into software-define, and in turn programmable with agility and optimization driving the trend. The software-define environment made possible the customizable tools for the workloads and mapping of these components onto the programmable infrastructure. Efficiency, resiliency, and security is an additional advantage\cite{50}.

Software-define potential enables decouple resource concept of traditional physical resources, e.g., memory and processor, which are tied to their hardware chassis through resource disaggregation. This takes into consideration physical resources as an individual component. These resources are then controlled by software. The software-defined provide flexibility and modularity to cloud infrastructure, which gives the opportunity of allocating new resources and applications due to the physical nature of the infrastructure. Software-define technology abstracts the infrastructure by making it programmable \cite{76}, giving the mobile operators the ability to upgrade part of their infrastructure component for better performance. Instead of changing the entire servers because each component has a different life-cycle, allowing the maximum degree of resource utilization.

Analysis in \cite{77} shows that disaggregated servers used in telecom networks minimize up to 50\% of the edge node network bottleneck compared to the traditional servers. This allows the segregation of traditional server resources into a pool of resources, proving high bandwidth and low latency networks. SDI can be used in designing decentralized big data governance activities \cite{78}, in which consumer applications producing data, such data can be decoupled from its application by making it independent of the application with implementation of rules and policies regards to applications producing data. In this fashion, data owners have full authority to decide where their data can be stored and how it can be shared. This paradigm decentralizes data through SDI in establishing and generating new inventions by linking data as well as best adaptation to an open environment.

Based on SDI, \cite{79} proposed a new paradigm that is SDS. Traditional data center faces challenges in implementing Quality of Service (QoS). SDS provides the solution, with an increase in the number of data center storage. The conventional method of data storage is unable to deal with the large amount of
data, methods of approaches like virtualization of data storage and algorithms for placement of data can be used for a fixed network topology of a data center. However, SDS can provide a solution, with requests of services that require a minimum amount of response time from data storage. The SDS abstract the storage control and implement a centralized control to the software layer. Infrastructure transformation is required by current applications and services in meeting customer requirements, as such experimental analyses are performed to show the advantage of SDS. Results of the experiment show that SDS and data placement algorithm achieve great performance compared to the physical storage, that is solid-state drive (SSD) and hard disk drive (HDD). The software-defines enable abstraction of network services, control, and management functionality \cite{72}. This approach allows mobile operators to select a range of services that can include fast deployment of new and enhance services, creation of a customizable network, and enhanced operation which can be extended to another software-define, such as cloud, storage devices, and many others, creating services than conventional or legacy network.

\begin{figure}[h]
	\begin{center}
		\includegraphics[width=3.5in]{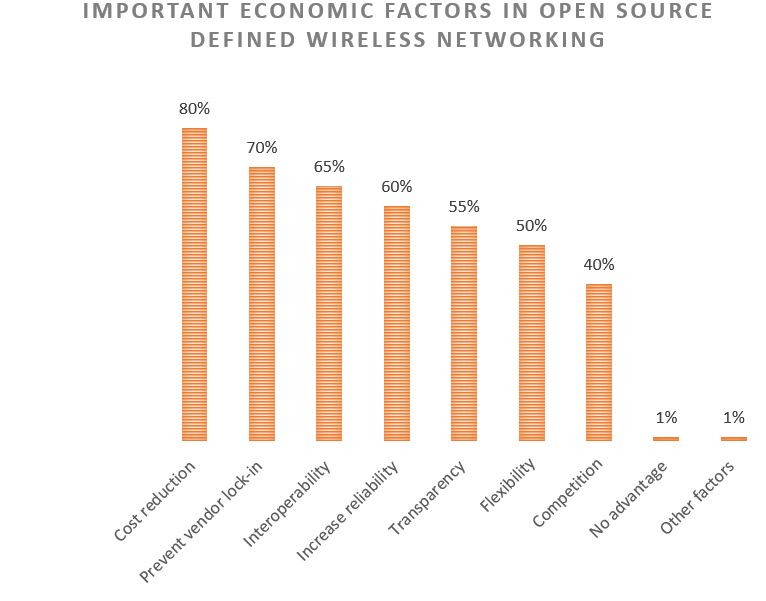}
		\caption{Analyze economic factor of open-source defined wireless network}
		\label{fig9}
	\end{center}
\end{figure}

This led to the proposal of a Universal Operating System (UOS) for SDI in \cite{73}, that will be a distributed operating system. The UOS will span between terminals (e.g., handset, tablet, smart-things, drones, etc.) through network elements, then to cloud/IT resources. This will have a great impact on the techno-economic of telecommunication and cloud/IT ecosystems. SDI will pave the way in executing any network function and services that are developed and deployed as a software application. This creates room for new operators to take part in the game, due to low CAPEX and OPEX, as well as a reduction in power consumption, which is optimized with the SDI paradigm. Hence, Fig. \ref{fig9} shows analyze factors economic importance of open-source-defined wireless networks.

Furthermore, Hu et al. \cite{019} propose Software-Defined Devices (SDD) from the operational mechanism of cyber-physical mapping perspective, and based on device-as-a-service an open IoT system architecture that decouples application at the upper-level from physical devices through SDD mechanism. With a designed logically centralized controller that manages the physical devices and the flexible device discovery, and device control interface with application request. In order to solve the problem of IoT interconnection and intercommunication of heterogeneous devices with improvement in utilization and efficiency in device resources, SDD-based open IoT system architecture has been implemented by the authors with modular customization of underlying IoT resource devices, such as unified management and scheduling, reusing, sharing and recombining of the devices mentioned above. 

Software-Define Edge Computing (SDEC) is proposed by\cite{a019} from the perspective of cyber-physical mapping with the aim of obtaining a high automatic and intelligent edge computing system. 
Through software-defined, SDEC enables flexible management with intelligent coordination amidst the various edge hardware resources and services. However, SDEC-based open IoT architecture decouples the topmost IoT application from the underlying physical edge resources and implement a dynamic re-configurable smart edge device. With the SDEC platform, detailed processes of the underlying physical devices can enable reuse, sharing, recombination, and reconfiguration of edge resources and services in improving the capacity of edge devices.   

\subsubsection{Programmable}

The need of making the network more scalable, programmable \cite{b019} robust with innovation of ideas and protocols result in the SDN concept. SDN architecture enables the network to become programmable through the interfaces and protocols mentioned below. It decouples the control plane from the data plane and centralized the controller by abstracting the lower-level infrastructure with a global view of the network. Initially, the telecom industry does not implement or embrace this concept. However, high CAPEX/OPEX and predicted data traffic from billions of devices connected to the cellular network, forces mobile operators in seeking a solution, and keep their customer's demands with better user experience and annual revenue. The programmability of SDN is discussed below from the perspective of northbound and southbound interfaces, and open API. 

\begin{figure}[h]
	\begin{center}
		\includegraphics[width=3.3in]{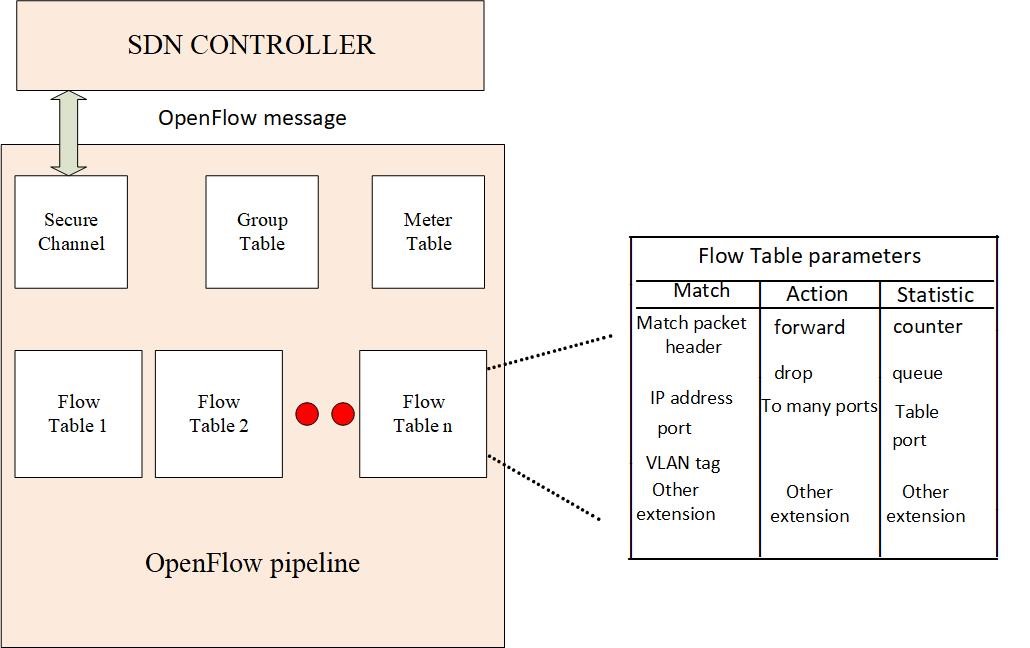}
		\caption{Structure of OpenFlow table}
		\label{fig10}
	\end{center}
\end{figure}

The evolution of the programmable open network is presented in the work of Anerousis et. al\cite{v}. The authors reviewed the origin of open and programmable networks and SDN, with numerous works on programmable networks that illustrate the emergence of successive SDN features from the concept of telephone network control in 1990. This led the way to the current effort of IEEE's standardization of network programmability with OpenFlow which influence the modern architectural transformation of next-generation networks.
\begin{itemize}
\item open-source on Northbound Interface (NI):  
This interface provides a high-level interaction of elements to implement network applications and services by programming the SDN controller (e.g., ONOS, ODL, etc.) via an open API. In this, REST API is used with HTTP in describing the Open Cloud Compute Interface (OCCI) used by application developers and offers interaction to the northbound interface that bridges the control plane and the management plane. REST design architecture provides interoperability, flexibility, and scalability to the northbound interface. The fact that northbound interface lacks standardization compared to the southbound interface, no single API that vendors or application developers can use. Though most controllers and application developers used REST API \cite{82}. 
REST API interacts with a core component that is responsible for distributing and positioning infrastructure through HTTP protocol. The combination of these components provides scalability in customizing the complicated IoT infrastructure. In fact, REST API has become ubiquitous in most distributed resources including the northbound interface due to dynamic upgrades from time to time, to prevent the client from crumbling. This can be achieved through the REST chart \cite{83}, which is a Petri-net service of REST framework that can be used in designing an extensible REST API to withstand any changes to the northbound interface. To achieved fast navigation with REST chart through a large number of REST APIs with reduced hypertext navigation overhead at the client-side, backtracking and cache mechanism can be applied to reduce the client-side overhead with the sustainable requirement of extensibility and flexibility \cite{84}.

\item open-source on Southbound Interface (SI):
On this interface, the lower-level protocols (e.g., controller communicate with other controllers) and the forwarding devices (i.e., switches) are programmed to communicate through a standardized and open protocol. 
The southbound interface solved the problem and challenges faced by heterogeneous vendor-proprietary network devices, by proving a common API used by different vendors. OpenFlow \cite{86} is the famous southbound interface protocol used in the SDN environment for communication between the controller and the infrastructure layer or data plane layer.  
OpenFlow \cite{87} evolve from version 1.0.0 through version 1.1.0, 1.2.0, 1.3.0, 1.4.0, and 1.5.1 respectively. Version 1.0.0 consist of a header, counters, and actions \cite{88}, and do not have much flexibility. As a result, switches that use this version cannot perform multiple operations at the time of packet forwarding because of one flow table and mixed matched field. 
However, to solve this problem OpenFlow version 1.1.0 \cite{87} introduce multiple tables through the use of pipeline processing.  OpenFlow switch is classified into two categories, the OpenFlow only and the hybrid OpenFlow. The former process all the packet in the OpenFlow pipeline and cannot be further processed otherwise. The latter can support both OpenFlow processing and Ethernet switch operation (e.g., conventional L2 switch, L3 router, etc.). This simply provides a traffic mechanism to route packets either through the OpenFlow pipeline or the usual pipeline. OpenFlow version 1.2.0 \cite{89} flow match field, used OpenFlow Extensible Match (OXM) in describing the match field which is a Type-Length-Value (TLV) format and OXM TLV range from 5 to 259 bytes. OpenFlow 1.2.0 introduces multiple controllers that improve reliability in case of failure of one controller. In this case, the switch will operate the OpenFlow routine if one controller connection fails. This multiple controller mechanism deals with only controller fail-over and load balancing. 
In addition, OpenFlow version 1.3.0 introduces the meter table consisting of meter entries, which enable the implementation of QoS and also provide the table miss entry. In this way, a packet that is unmatched by other flow entries is sent back to the controller. OpenFlow version 1.4.0 introduce a synchronization table, with two flow table that can simultaneously be synchronized, resulting in automatic update by the switch to reflect the flow table which is synchronized. A flow table can be synchronized bidirectional or unidirectional. Bidirectional synchronization reflects changes made by the controller to the source switch table. For any synchronized table, a particular table property can describe the source flow table from which it synchronizes. Flag extension was introduced in OpenFlow version 1.5.0 on the flow table. The flag extension alters the pattern in which flow entries are managed. For example, OFPP SEND FLOW REM, this flag removes the message from that particular flow entry. Fig. \ref{fig10} represent the diagrammatic structure of the OpenFlow table, which gives the details specifications on how to transform or migrate the control logic of a switch into the controller as well as defining the protocol of communication between the switch and the controller \cite{90}.

\end{itemize}

It is a software-based controller that manages the forwarding piece of information in one or more switches. While the hardware takes control of forwarding the traffic based on rules from the controller. In the field of innovation, OpenFlow architecture provides the capability for researchers to experiment, test, and initiate new ideas or applications. The capabilities include dynamic update forwarding rules by the controller, centralized control and traffic analysis \cite{91}. This means that if the control plane is managed by a hardware device, experiments and research will be chaotic due to the firmware modification difficulty. Application base OpenFlow proposed by \cite{92} make it easier for network configuration, simplicity of network management, the addition of security features with respect to virtualize network, and deployment of mobile system. In a nutshell, OpenFlow gives the details specifications on how to transform or migrate the control logic of a switch into the controller as well as defining the protocol of communication between the switch and the controller \cite{93}. In table \ref{tab.V}, we present a summary of OpenFlow enable devices.

\begin{table*}[]
	\caption{Openflow enable devices}
	\label{tab.V}
	\centering
	\begin{tabular}{|l|l|l|l|l|l|}
		\hline
		\bfseries Classification            & \bfseries Product name                   & \bfseries Product type    & \bfseries Designer           & \bfseries Description                                                                & \bfseries Ref.     \\ \hline
		\multirow{5}{*}{Hardware} & Pica8 3920                                                     & Ethernet switch & Pica8              & \begin{tabular}[c]{@{}l@{}}Built for open switch and \\ perfect for virtualize data center\end{tabular}      & {}\cite{94}{} \\ \cline{2-6} 
		& CX600                                                          & Router          & Huawei             & \begin{tabular}[c]{@{}l@{}}Provide solution for carrier-class\\ metropolitan area network (MAN)\end{tabular} & {}\cite{95}{} \\ \cline{2-6} 
		& \begin{tabular}[c]{@{}l@{}}RackSwitch\\     G8264\end{tabular} & Switch          & IBM                & \begin{tabular}[c]{@{}l@{}}Switch for data center \\ that can support OpenFlow\end{tabular}                  & {}\cite{96}{} \\ \cline{2-6} 
		& Cisco catalyst 2960X/XR                                        & Switch          & Cisco              & \begin{tabular}[c]{@{}l@{}}Support for OpenFlow controller on catalyst \\ 2960X/XR\end{tabular}              & {}\cite{97}{} \\ \cline{2-6} 
		& V580                                                           & Switch          & Centec network     & \begin{tabular}[c]{@{}l@{}}Ethernet silicon CTC8096\\ leveraging OVS\end{tabular}                            & {}\cite{95}{} \\ \hline
		\multirow{4}{*}{Software} & Open vSwitch                                                   & Switch          & Open community     & \begin{tabular}[c]{@{}l@{}}Design to enable massive network \\ automation through programming.\end{tabular}  & {}\cite{98}{} \\ \cline{2-6} 
		& Indigo vSwitch                                                 & Switch          & Linux              & \begin{tabular}[c]{@{}l@{}}open-source OpenFlow that uses \\ Linux kernel forwarding module\end{tabular}     & {}\cite{99}{} \\ \cline{2-6} 
		& Big switch                                                     & Big vSwitch     & Big switch network & \begin{tabular}[c]{@{}l@{}}Network virtualization incorporate \\ open SDN architecture\end{tabular}          & {}\cite{56}{} \\ \cline{2-6} 
		& EX4600                                                         & Switch          & Juniper networks   & Support OpenFlow version 1.0.0 and 1.3.1                                                                     & {}\cite{100} \\ \hline
	\end{tabular}
\end{table*}

\subsection{MultiFire-Based}
The increase in voice and data traffic has led mobile operators to consider the deployment of LTE-like services in the unlicensed spectrum to fulfill consumer demand. This deployment of standalone LTE radio technology in the 5 GHZ unlicensed spectrum can further enable business opportunities, through leveraging the LTE-like features. MultiFire enables the deployment of standalone LTE radio technology, based on two candidate technologies proposed by 3GPP\cite{a93} (i.e., LTE enhances licensed assisted access and LTE-unlicensed). The coexistence of other technology, such as WiFi which used the technique of Listen Before Talk (LBT) is similar to Carrier Sense Multiple Access with Collision Avoidance (CSMA/CA) in the WiFi domain. This is an LBT-based MAC protocol that enables LTE-unlicensed to access the channel when it is idle, following a clear assessment called the backoff period. However, alternative channel medium access in time-division multiplexing (TDM) based MAC protocol called duty-cycle LTE. MultiFire extends the LTE protocol stack to implement flexible resource allocation and control of power consumption. Precisely, MultiFire targeted goal is to enhance Mobile BroadBand (eMBB) as well as IoT applications capable of supporting largely connected devices\cite{b93}. This extends the operation of LTE in licensed spectrum to cover the unlicensed spectrum.

In an effort to demonstrate the coexistence harmony between LTE and WiFi in the unlicensed band,\cite{c93} propose an Inter-Cell Interference Coordination (ICIC) based spectrum mechanism extending into an auction framework for efficient spectrum sharing in LTE-unlicensed band. Through Carrier Sensing Adaptive Transmission (CSAT) that led to LBT enhanced CSAT scheme. Furthermore, the work in\cite{d93} studies the energy-based CSAT in LTE-unlicensed (LTE-U). With the LTE-U protocol base station, allowed to use a free channel for 20ms and turn off its transmission for 1ms, this prevents another access point from using the same channel medium since the interval is low. This motivated the work in\cite{d93} to propose an LTE-U base station transmission scheme, not exceeding the maximum transmission rate of 20ms in a dense deployment of LTE-U and WiFi, even if the channel medium is empty for another wireless access point (e.g., WiFi) coexistence in the same channel.

\subsection{Distributed Management}

In this subsection, we discuss the open CN distributed management which enabled the integration of different SDN controllers into the CN with a global view of the network, that realizes open network decouple.  
In a conventional network, the lack of a centralized controller results in not providing a wide network management abstraction. This resulted in each novice network function (e.g., routing) producing its own particular state distribution, failure recovery mechanism, and element discovery. However, the absence of not having a unique control plane platform prevents further development of reliability and flexibility of the network control plane. To solve this problem, a distributed controller platform that is open, and vendor-neutral platform to network control plane can be used as the distributed system. This control plane operates with a global view of the network and provides a unique API for control plane execution \cite{80}. The result in decoupling of the control plane from the forwarding plane (i.e., user plane), resulted in a distributed system. Here the control plane handles the state distribution by taking information from the switch and distributing the control state through organizing the state in the server's platform. An example of this distributed open-source controller is the ONOS, which is distributed, logically centralized, with a global view of network, scalability, and fault tolerance. This also includes traffic engineering as well as scheduling. 

To cope with the trend of evolving technology, the study in \cite{81} introduces the concept of 5G with ONOS. Taking into consideration SDN/NFV’s potential role in the telecommunication domain. SDN and virtualization are good candidates in 5G, as wireless and mobile networks continue to be in significant demand.
In the work of Sambo \cite{i} delegated restoration of SDN disaggregated network controller is presented, which implemented a hybrid centralized distributed fashion. The author's motivation to innovate the novice paradigm result from the YANG data model that describes the vendor neutrality of network devices (i.e., switches and transponders) lacks the descriptive function. Whereby the control plane centralization can be affected in the process of scalability, such as link failure due to multiple restoration requests occurring at the same time. However, the SDN controller computes centrally the backup light path prior to a failure through communication with the devices (i.e., switches and transponders), via NETCONF in the case of link failure. Result analysis through simulation shows delegated restoration is able to reduce restoration time of a fully centralized approach to achieved operation, administration, and maintenance of next-generation networks.

The scalability of SDN networks on a single controller cannot meet the required network traffic, rather than having a centralized logical but physically distributed controller architecture which enhances the reliability and scalability of the controllers. This ideology however consists of controller load imbalance that the authors in \cite{ii} propose distribution decision mechanism (DDM), which is based on switch migration within large subdomain SDN network to address this problem. Through the network information distributed migration, the decision is created based on the controller load situation. Based on switch migration decision targeted controller is selected in probability order and integrated three network properties which include, switch migration, data collection, and controller state synchronization. This proposed mechanism achieved controller load balancing with good performance.

Almadani et. al \cite{iii} propose Distributed SDN control plane Framework (DSF) for heterogeneous west/east interface. These distributed SDN controllers use synchronization topologies with standard data-centric Real-Time Publish/Subscribe (RTPS) model, referred to as Data Distributed Service (DDS). The author's work was motivated by distributed controllers SDN west/east interface lacking communication standard for heterogenous, multi-domain network ecosystem. DSF framework control plane elements are arranged in a fashionable manner of vertical (i.e., child/parent) and horizontal (i.e., peer-to-peer), manner with Link Update Message Protocol (LUMP), to synchronize the topologies implemented by routing data plane packets across multiple domains. However, the increase in the number of controllers’ topology synchronization arises in real-time with the controller optimization platform selected. Togou et. al \cite{iv} propose HiDCoP, which is a high-performance distributed path computation in a large-scale SDN control plane with a hierarchical design to distribute the path computational load between the controllers, to minimize the transmission overhead. With 3-tiers architectural network domain, which is further subdivided into areas coordinated and maintained via topology information from controllers to accelerate the computational process, through parallelism and load balancing. HiDCoP simulation result proves the out performance that exists in path computation, path setup latency, and end-to-end delay. 

However, assessing the maturity of SDN controller is an utmost important factor to operators before deployment into the operational environment due to bugs in SDN software controllers, resulting from heterogeneous complexity of the network and forwarding devices the controller support. Vizarreta et al \cite{018} focus on Software Reliability Growth Models (SRGM), which model the stochastic nature of bugs manifestation in process of open-source SDN controllers. This enables network operators to determine the maturity of controller software to be deployed into the operational environment, based on network application requirements. SRGM can predict the stochastic analysis of bugs manifestation and correction procedure with real data on software failure in the SDN controller. Therefore, predicting the controller reliability. Hence, this provides the guide to network operators and software developers in tackling operations that are important in decision making, such as the prediction of time in releasing and deploying a matured SDN controller.

\begin{figure*}[]
	\begin{center}
		\includegraphics[width=6.3in]{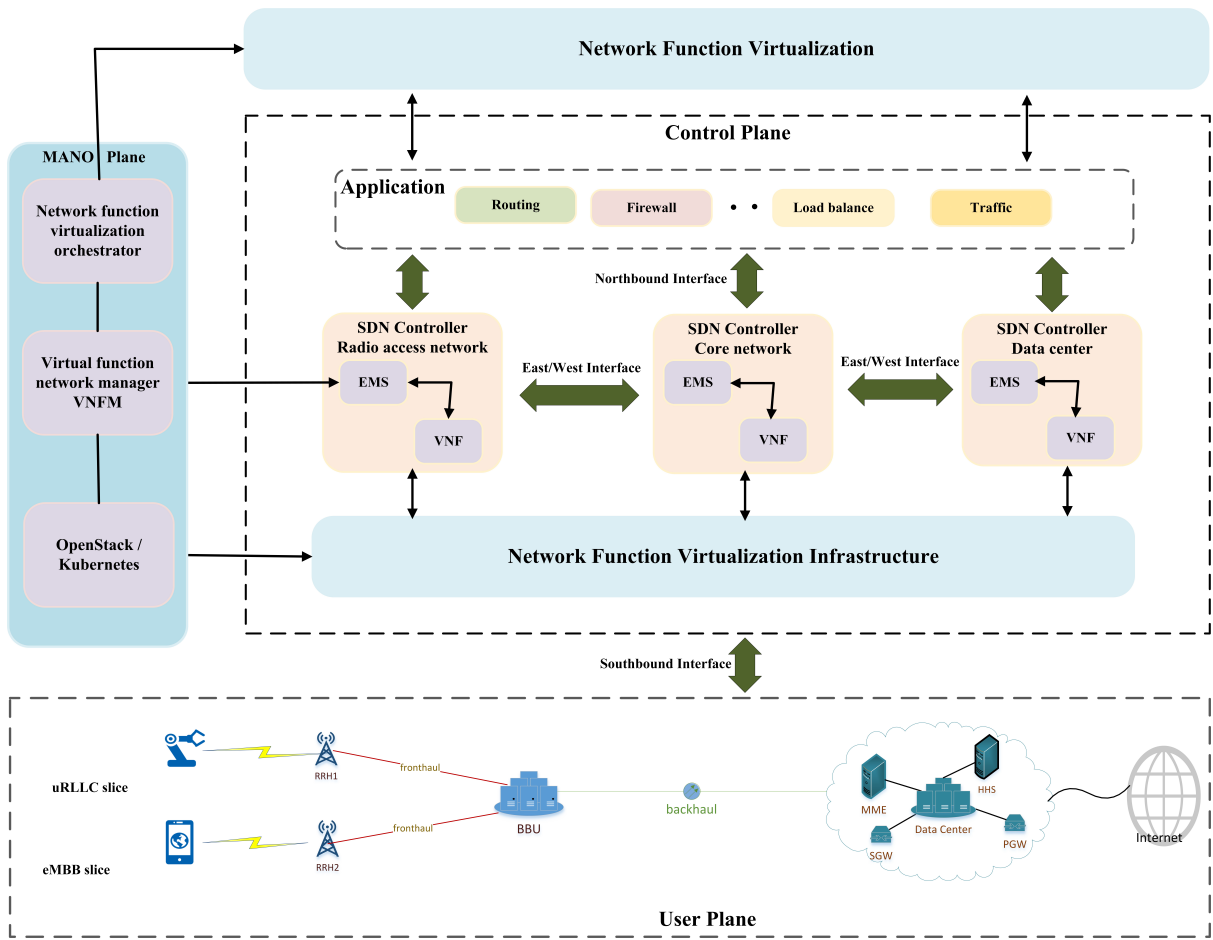}
		\caption{Conceptual ideology of federated SDN controllers based open-source-defined wireless networks}
		\label{fig11}
	\end{center}
\end{figure*}

In the context of open-source-defined wireless networks, Fig. \ref{fig11} represents the conceptual ideology of distributed SDN controllers but logically centralizes at the control plane. 
The NFVI consist of an abstracted logical environment that serves as the building block (e.g., compute, storage, and network) resources. In addition, NFV in the context of network slicing provides the potential framework that implements the service-chaining of VNFs. Management and orchestration (i.e., MANO plane) control, orchestrate, manage, and coordinate VNFs and NFVI. With the network resource orchestration through NFV Orchestration (i.e., NFVO), the VIM validates and authorizes the NFVI resource requested from the VNF manager. Furthermore, in a service chain of multiple VNFs Element Management System (EMS) orchestrate and manage VNFs life-cycle through the VNF manager.

The choice of controller placement is an important role in the network's latency and resiliency. As such the controller placement problem can decide the number of controllers required in the network. The work in \cite{x016} surveyed the controller placement problem in SDN network, based classical formulation that support system model and associated metrics with objective and methodology. Controller placement problem impact on the novice application domain, such as mobile network, named data network, and wireless mesh network cannot be ignored due to the energy efficiency in load balancing.  

\section{Open-source-defined Radio Access Network}\label{AA}
In this section, we discuss open-source-defined RAN which is the key to our survey-based disaggregation of the RAN hardware and functional components. A lot of effort has been made in RAN virtualization through open-source software tools to substitute the legacy telecommunication protocol, and conventional proprietary complex system \cite{101}. For example, in 2010, the OpenBTS (Open Base Transceiver Station) \cite{102} which is open-source software for GSM (Global System for Mobile) was presented. OpenBTS provide the opportunity to develop more than one carrier signal on the same physical radio, in order to increase the network capacity. In addition, Software-Defined Radio (SDR) is the game-changer that enables OpenBTS to become possible through the implementation of a Radio Frequency (RF) between 65 MHz to 4 GHz, attached to a host application, such as FM Radio. SDR made it possible for the implementation of open-source RAN as software on a commodity hardware processor. For example, open-source full-stack IEEE 802.11a/g/n SDR implementation is used to demonstrate OpenWiFi, based on System-on-Chip (SoC) and include Field Programmable Gate Arrays (FPGA) and ARM processor. With the implementation of drivers embedded in Linux OS which run on ARM processor. These drivers are used in instantiating the APIs defined by Linux OS, which allows researchers to study and modify OpenWiFi for its modular design.

However, FPGA low latency connection with RF front-end promotes the implementation of the physical layer (PHY), and lower Media Access Control (Low MAC) in FPGA \cite{169}. 
To achieve low latency at both PHY and MAC, a new SDR ``Tick" is proposed which is modular in design and programmable, validated by the authors through extensive evaluation and prototyping of 108.11ac SISO/MIMO and 802.11a/g full-duplex \cite{170}. 
Bloessl et al. \cite{031} present both simulation and experimentation framework of IEEE 802.11 p based SDR Orthogonal Frequency Division Multiplexing (OFDM) with extensive validation test for interoperability, field trial, and important benefits obtained through SDR-based transceiver. The transceiver's transparency open and programmable hardware and software, implement and provide detail access to all data below, including the physical layer for better understanding of the system. This provides seamless switching between experimentation and simulation that bridge the gap between theory and practice and serve as a tool for further study of IEEE 802.11 p in both simulation-based and field operational tests for novel physical layer solution. Considering the fact that both IEEE 802.11 a/g and IEEE 802.11 p share similarity in physical layer-based OFDM, with IEEE 802.11 p having double-timing.

OpenBSC (Open Base Station Controller) open-source software was launched to provide a base for experimentation with GSM in academia as well as research centers \cite{103}. SrsLTE \cite{104} is an open-source platform for LTE experimentation with the open-source library for the physical layer of LTE release 8, that has maximum modularity of code reused which is written in ANSI C. SrsLTE is implemented with Single Instruction Multiple Data (SIMD) operation, and provide the interface for Universal Hardware Driver (UHD) enabling the support for Ettus USRP (Universal Software Radio Platform) devices. In fact, the goal of the library is to provide the necessary tools for the implementation of LTE-based applications, e.g., eNB, UE, and so on. 

OpenAirInterface (OAI) is a 4G/LTE open-source software \cite{105}. Inline recently with OAI 5G that can implement (eNB, UE, CN) from a general-purpose hardware, integrating SDN/NFV and OpenStack to established efficiency in RAN. OAI Open-source project main goal purpose is softwarization of mobile network functions implementation of 3GPP technology, ranging from network access, e.g. OAI-RAN to core mobile network, i.e., OAI-CN used in 4G mobile network prototyping. This enable implementation on general purpose hardware, e.g., x86 architecture. OAI open-source project offers software implementation of all components, such as UE, eNodeB as mention above via software radio front end connected to the host computer which enables innovation in the mobile network communication field \cite{14}. OAI 5G NR implementation is discuss in \cite{14a}. The authors implemented the basic functions required to support downlink functionality at both gNB and UE to realize 5G NR in real-time on software defined radio platform and compliant inter-operable with other commercial equipment. The joint effort R. -G. Cheng et al. \cite{14b} demonstrate the design and implementation of open-source narrow band IoT based OAI to enable services in licensed spectrum, which enable an operator to replace GSM carrier in stand-alone mode by adopting the orthogonal frequency division multiple access technology. Based on OAI, new network functional application platform interface (NFAPI) is defined to support virtualized MAC/PHY split for 5G.     

SrsLTE open-source platform provides LTE experimentation designed and code reused modularity compliant with LTE release 8. In \cite{104} present the SrsLTE library as well as SrsUE which extends and implements LTE transmission in unlicensed bands with the coexistence of WiFi. In release 10, SrsLTE enables the implementation of LTE eNB, UE, and EPC that support compatibility with Linux. Several research effort has implemented srsLTE in various research areas. Inter-slice RAN controller-based srsLTE is implemented in \cite{14c} to effectively manage radio resources for different slices. LTE control plane security attack is demonstrated in \cite{14d} to discover, design, and implement vulnerabilities resulting from carriers and device vendors. 

However, the initial motivation to shift focus on RAN disaggregation is to have ultra-low latency and high radio speed, with a combination of the massive amount of connectivity in 5G, which the fronthaul interface of C-RAN cannot withstand. This resulted in industrial forum, such as O-RAN and TIP to focus on decoupling the RAN control plane from the user plane, resulting in the modularity of the RAN into a stack that can run on commodity hardware with both open north-southbound interfaces. The concept of network slicing bridges the disaggregated RAN component closer to softwarization and cloudification of decoupled NFs to realize a flexible virtual RAN (vRAN). 
With the ongoing evolution of open-source taking part in current and future technology. Network cloudification \cite{121} comes into play role. This consists of SDN/NFV, Mobile Edge Computing (MEC), as well as centralizing and virtualization of the network in the mobile operator's data center. This makes it possible for operators to provide Network-as-a-Service (NaaS) at a lower amount of CAPEX and OPEX, resulting in simplification of networking devices and providing easier orchestration of services. To showcase MEC service placement in disaggregated heterogeneous C-RAN with both cellular and WiFi infrastructure to provide access fronthaul network resources in multi-home-used UEs is demonstrated in \cite{036}. Lower latency is achieved through the selection of a possible wireless network at the MEC side.
\subsection{O-RAN Approach} 
In this subsection, the splitting of the RAN hardware, software, and functional components from their tightly couple dedicated hardware is discussed. The decouple software with open specifications running on the commodity hardware permits operators to invent their applications faster with reduce cost of operation. 
\subsubsection{Splitting of Functional Component}
In the 5G transport system, transport options require a rethink solution of the functional split for RAN. However, in 5G the baseband unit is then split into CU, DU, and RRU as defined by WG8 and WG6 respectively. The redesign facilitate RAN virtualization in addition to reducing the fronthaul capacity as well as latency requirement. The CU is part of the next generation NodeB (gNB), which handles user data, radio access network sharing, session management, mobility control, etc. Moreover, CU coordinates the controls of DUs over a fronthaul interface. The DU is another sub-part of the gNB as part of the 3GPP functional split. This function split improves the RAN flexibility and minimizes the infrastructural deployment cost without affecting the user QoS. In view of the above split, Matoussi et al. \cite{020} propose a novel functional split orchestration scheme based on minimization of RAN deployment cost ``AgileRAN" that implements cloud capability of on-demand deployment of RAN resources.

AgileRAN virtualizes and splits the baseband processing chain through a heuristic-based approach. This takes into consideration the processing network function requirement, with the capacity of cloud infrastructure. In addition to this novel functional split, both processing and bandwidth resource usage is optimized at the same time minimizing the energy consumption, through 5G experimental prototyping on the OAI platform good performance results in total deployment cost and resolution time are obtained. The disaggregation of the functional component has opened a research ground for both academia and industries. 
This functional split is taken LTE protocol into consideration with eight possible options for the split by 3GPP. In this split, we will map each split with the appropriate O-RAN WG approach for more clarity.

\begin{itemize}
	\item Option 1 / WG 8  (RRC / PDCP split):
	This split user plane (UP) is located at the DU giving the user data benefits proximity to the transmission point which enables caching. However, in a CU-pool connected to many cells other research work has shown that it is not of beneficial importance due to fact that it does not support several functions to implement inter-cell coordination \cite{047}.
	\item Option 2 / WG 6  (PDCP/ RLC split): 
	Both PDCP and RLC functions are centralized whereas another function resides locally in the DU. Within this split, the traffic is subdivided into multiple flows that provide multiple directions to various access nodes which support multi-connectivity. 
	\item Option 3 / WG 5 \& WG 3 (High  RLC/Low RLC split, intra RLC split:)  
	This split consists of the high RLC and low RLC with low RLC consisting of the segmental function and high RLC consist of automatic repeat request (ARQ). Both the user plane (UP) and PDCP RLC asynchronous processing occur at the CU, with remaining UP function residing in the DU. This WGS implement the open interfaces, which include F1, W1, E1, and E2, as well as RIC near-Real Time respectively. 
	\item Option 4 / WG 6 (RLC MAC split):
	In this optional functional split, the RLC in the downlink (DL) receives the protocol data unit (PDU) and channel MAC service data unit in the uplink (UP). This split when virtualized can enhance the benefits of resource sharing for the utilization of both storage and processing capability. 
	\item Option 8 / WG 7 (PHY-RF split): 
	In this functional split, both the RF and up-converter are left at the DU which as a result support different radio access technology (RAT), whereas the rest of the functions are centralize\cite{113}.
	However, the use case scenario and RIC non-Real Time-base A1 open interface are focused on WG 1 and WG 2 respectively. 
\end{itemize}

\subsection{Open5G Approach}
In this subsection, we discuss from the Open5G open-source community perspective open RAN, key enabling technology, orchestration, and network slicing of decoupling software. However, the open wireless network-based 3GPP disaggregation of RAN into different components and open interfaces, such as CU/DU, and open-source software. In this overview, the CU implements the RAN-NFVI and extends the concept of SDN. Precisely the CU-control plane and CU-user plane with RRC/PDCP-control and PDCP-user plane CU-protocol stack, open interfaces, and RIC for intelligent decision making. On the other hand, Open5G open-source community approach envisioned DU from the physical high layer (i.e., RLC/MAC) and x86/x64 architectural general-purpose platform interfacing both F1 to the CU and NFGI-1 to the physical lower layer of white-box RRU \cite{i006}.

In addition, the concept of network slicing realizes the disaggregated RAN component functions into softwarization and cloudification of VNF. In this context, network slicing integrates or unified the virtual resources that include VM with the instantiation of VNFs which are linked together via a virtual network (e.g., virtual Local Area Network or Virtual Private Network) to provide a service of multi-VNFs called service chain \cite{043}. This can be achieved with the enabling technology of network slicing, such as SDN/NFV, hypervisor, cloud \& edge computing, container and VM.

\begin{figure*}[t]
	\begin{center}
		\includegraphics[width=6.4in]{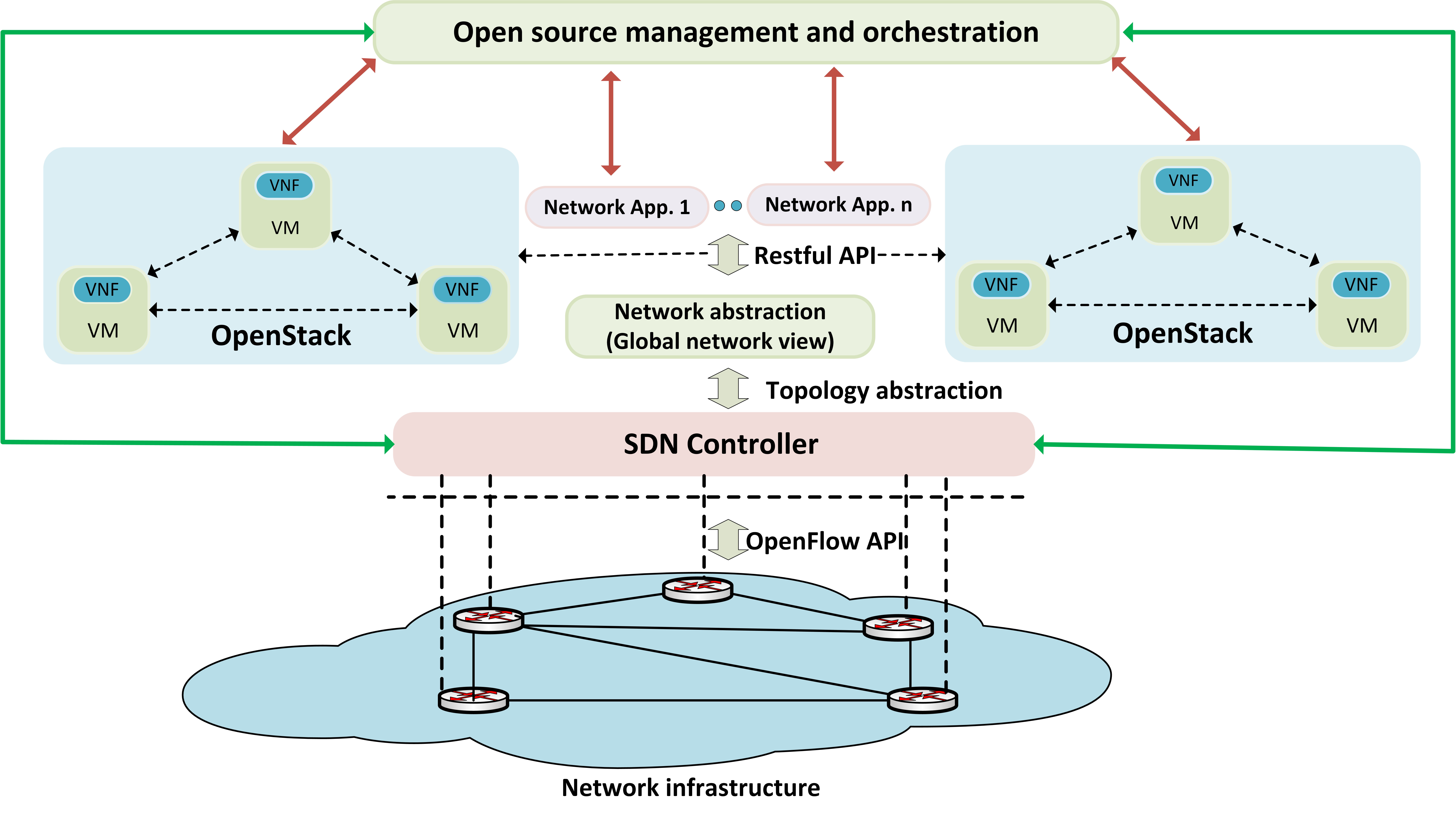}
		\caption{The abstraction of SDN and NFV interfaces.}
		\label{fig12}
	\end{center}
\end{figure*}

The proof-of-concept framework leverage an end-to-end OSS base mobile network as discussed in Section \ref{BB}-A, integrate prioritization algorithm technique to achieve higher throughput and lower delay for time-sensitive and important services, such as connected cars which can enable RAN slicing of end-to-end network. However, open-source control management and orchestration of NFs in SDN/NFV enable networks through programming, with third-party application access to the northbound interface of SDN in the realization of Optimization-as-a-Service (OaaS) presented by \cite{108}. In addition, major OSS exits in the network resource allocation process and heuristic technique in optimization with solvers (which are important libraries used in figuring out numerical solutions to particular structured optimization problems in a structure linear or nonlinear formulations) in SDN/NFV environment.

The OpenStack \cite{116} controls a pool of resources, such as compute, storage, and networking resources in the data center, managed from the dashboard or from the OpenStack API. This software is popular among enterprises and open-source technologies which makes it a perfect ideal for heterogeneous network infrastructure. However, Nova \cite{117} project provide an open API service implementation, massively scalable self-service, and access to compute resources including VMs, bare metal, and containers.
open-source NFV MANO \cite{118} with ONAP\cite{119}, enable automation, management, and reconfiguration of VNFs deployed into the cloud. ONAP provides a detailed comprehensive platform for orchestration and automation with physical and virtual network functions, that allows cloud providers, networks, developers to develop new services and sustain the complete life cycle management. With the open standard, ONAP can fasten the development of a vibrant ecosystem by unifying resources. In table \ref{tab.VI}, we represent the difference in open-source between pre-SDN/NFV and post-SDN/NFV. 
\begin{table*}[t]
	\caption{The difference with respect to open-source between sdn and nfv}
	\label{tab.VI}
	\centering
	\begin{tabular}{|c|c|c|}
		\hline
		\textbf{Category}	& \textbf{Pre-SDN and NFV}                                                                                & \textbf{Post-SDN and NFV}                                                                                                              \\ \hline
		Application area                   & \begin{tabular}[c]{@{}c@{}}Mostly used in particular parts to provide a solution\\ E.g. OpenSSL that is used as encryption area\end{tabular} & \begin{tabular}[c]{@{}c@{}}Mostly across all component, data plane, \\ control plane, management plane, and infrastructure\end{tabular}\\ \hline
		Device level used                  & \begin{tabular}[c]{@{}c@{}}open-source is used mainly in communication\\ libraries, and part of operating system\end{tabular}                & \begin{tabular}[c]{@{}c@{}}Used in full SDN controller providing useful\\ features and applications\end{tabular}              \\\hline
		Management and Orchestration level & Legacy and proprietary software used                                                                                                         & \begin{tabular}[c]{@{}c@{}}Provide a complete stack of open-source \\ solution like OpenStack\end{tabular}                       \\\hline
	\end{tabular}
\end{table*}

\subsection{Open Centralize Unit (CU) and Distributed Unit (DU)}
In this subsection, we focus on the open CU/DU. The open interface fronthaul Next Generation Fronthaul Interface (NGFI) between the DU and RRU of the RAN, permits smaller vendors and operators to take part and introduce services to their customers. Open interfaces allow interoperability between multi-vendor equipment and flexibility \cite{126}. However, the open interface in the RAN stems from the idea of disaggregation of functional component (i.e., RU, CU, and DU) in achieving openness in the control plane, user plane, and the management plane. The functional split option 7-2x \cite{130} at the physical layer between open DU and open CU bridge the gap of conventional fronthaul interface used by the previous generation, such as LTE fronthaul in which each vendor defines its own fronthaul specification which created a ``black-box", to a more interoperable multi-vendor or vendor-independent fronthaul interface. However, the higher layer split between PDCP and RLC, as well as the physical lower layer split, create advantage for operators to deploy virtualization infrastructure, such as RAN NFVI and VIM which leverage OSS in support of the functional split for virtualization. This split determines the amount of functions from the base station that are to be left close to the users, and which functions are to be centralized in the open CU for achieving maximum processing \cite{113}, and enable the evolve Common Public Radio Interface (eCPRI) to be implemented for traffic aggregation. 
\subsection{White-box Remote Radio Unit}
White-box Remote Radio Unit/Active Antenna Head (RRU/AAH) enables multiple technologies to be implemented, maintain, and run simultaneously in the same RRU. This reduces design cost to the operators and enabled continuous evolution of technology from 5G to 6G. Considering RAN split option 7, RRU can communicate to a separate server for authentication as discussed in section \ref{BB}-A. However, this split enables the implementation of Commercial-Off-The-Shelf (COTS) hardware with hardware custom accelerators such as programmable FGPAs. We summarize some important open-source software in the open networking ecosystem in table \ref{tab.VII}.
\begin{table*}[t]
	\caption{Important open-source software and hardware in the networking ecosystem}
	\label{tab.VII}
	\centering
	\begin{tabular}{|c|c|c|}
		\hline
		\textbf{Category}    & \textbf{Types}              & \textbf{Software example}     \\ \hline
		Control plane        & SDN controllers, L2 stack   & ODL, ONOS, Ryu, Open Contrail \\ \hline
		Management plane     & MANO                        & ONAP, OSM                     \\ \hline
		Infrastructure       & VIM, NFVI                   & OpenStack, OPNFV              \\ \hline
		Hardware abstraction & Networking operating system & ONL, ONIE                     \\ \hline
		Data plane L2-3      & Switches                    & OpenSwitch, OVS, OVN          \\ \hline
		Hardware design      & White box switches          & OCP, TIP                      \\ \hline
	\end{tabular}
\end{table*} 

\section{Open-Source Multi-access Edge Computing}\label{EE} 
In this section, we discuss the open-source Multi-access Edge Computing scheme which is based on splitting of conventional MEC NFs via NFV technology, and reconfiguration of traditional MEC elements as well as the extension of SBA concept into traditional MEC to realize open-source MEC. These independent NFs are supported by SBI and extending the MANO plane to MEC's NFs reconfiguration, with an Open-source MEC template instantiate when required. However, we will give a glance view of OpenFog as well to converge a clear picture. 
\subsection{Open-Source MEC Framework}
In this subsection, we discuss the open-source MEC framework which extends the SBA concept to realize the service-based MEC layer with the implementation of new NFs \cite{113a}. Similar ideology extension from the 5G core, these NFs can communicate via the SBI-based MEC layer with the web communication protocol i.e., HyperText Transfer Protocol (HTTP). However, to achieve the capability of 5G NR at the user plane. The extension of UPF is realized to enable open-source MEC to achieve 5G NR potential. The open-source MEC framework is designed with components layers which include infrastructure, virtualization, MANO, application, and the VIM that manage the infrastructure resources as discussed in Section \ref{BB}-A for more details. Open-source MEC framework decouples of NFs is achieved by the service-based MEC layer. The SBI initiate and established a connection between the NFs with programming capability. Furthermore, open-source MEC reconfiguration is enabled via templates and instances with Kubernetes \cite{48}. The open-source MEC template enables the pre-definition and selection of appropriate NFs for a particular application type. The open-source MEC instance enables edge resources/services in close proximity to the users via the instantiation process of a service request by the users.
\subsection{OpenFog} 
OpenFog was established by ARM, Cisco, Microsoft, Intel, Dell, and Princeton University to resolve the problem of bandwidth, latency, and communication challenges with IoT as well as related domains. This enables the processing of data information and intelligent at the edge. In this respect, many researchers have joint efforts in posturing novel ideology to improve the OpenFog architecture. To showcase the work in \cite{113b} propose a converge model to extend the ETSI MANO into OpenFog architecture which enable the combined capability for the IoT domain. The authors address how the combined implementation can bridge the gap between operational technology and information technology. This permits the automated and homogeneous orchestration of cloud node, edge, fog, applications, and interfaces with end-to-end security. This create a multi-tier communication and computing infrastructure domain resulting in new security challenges. As such the work in \cite{113c} present an overview security landscape of OpenFog architecture to enable trusted information services to various information and operational technology application. However, by leveraging NFV technology security-as-a-service can be deployed at the Fog node to protect the end devices against security threats.  

The OpenFog Consortium defined three-layer architecture i.e., service, communication, and security application layer to permit interoperability for security solutions in OpenFog. Reliable and secure communication is established via the communication secure layer for confidentiality, integrity, and authentication in data and traffic flow between node-to-node, node-to-cloud, and node-to-end-point communication \cite{113d}. To achieve resiliency in the network, the work in \cite{113e} propose the framework integration of fog node and SDN OpenFlow switch to enable inspection of data packets traversing the network. This relays messages to the controller for flow rules installation across the network, with IP sniffing application used to validate the framework.
\section{Security challenges, current and future evolution}\label{FF}
In this section, we present a review focusing on security challenges, current and future evolution regarding ongoing research efforts in open-source-defined wireless networks in current and future technology. With new paradigm, new security threat can be introduce. For example, the integrity and confidentiality of data information in VNF depend on the isolation property of the hypervisor which in turn depend on the integrity of mobile operator's cloud software \cite{131}. Every decade experience a new service and technology introduce into communication systems. Just like new services and use cases are been introduce in 5G, such as M2M communication, autonomous system, uRLLC and so on. The next generation wireless network (6G) will be accomplished with exited new features, such as fully AI, optical wireless communication, terahertz communication, three-dimensional communication (3D), quantum communication and so on \cite{132}, \cite{133}. Open-source-defined wireless network will play a vital role as well. 
\subsection{Security Challenges}
In this subsection, we discuss the importance of security with the current breaching of confidential data, security breached can have severe consequences. Hence, mobile operators’ data integrity is of utmost importance in protecting users’ data privacy. The security issue is an important aspect especially with regards to a centralized controller. An attacker can simply target the controller which acts as the brain of the network with attacks, such as Denial of Service (DoS), Distributed Denial of Service (DDoS), eavesdropping, identity spoofing, man in the middle attack, etc. Issues about back-door in SDN future mobile network are also addressed and solutions are provided by \cite{124}. Experimentation with ONOS shows its capable of offering beneficial used cases in network security\cite{a133}.

In the work of Lina and Dongzhao \cite{134}, they propose Software-defined Security (SDS) which is open and universal that provide security services and security management to security devices. SDS can enable different network security vendors in the deployment of security solutions and security products, by the realization of Virtual Security Functions (VSFs) through arrangement and customization. The authors analyze the different types of attacks and how the defender can disable the network attack by changing the security configuration of the server. TeFENS \cite{135} and Murthy et. al \cite{136} propose a security testbed for the next-generation mobile networks that enable the researcher to perform various security analyses, which is transparent and easy to deploy. This allows the researcher to conduct cyber-attack and control measures on how the defense measures are performed.

The author's testbed leverage some of the open-source project, such as open-source MANO, ONOS, and OpenStack in order to conduct stringent cyber-security analysis on different requirement of the network. Shang et al. \cite{142} present FloodDefender which is an efficient protocol-independent defense framework, for the OpenFlow network to detect and mitigate SDN-aim DoS attacks. FloodDefender fits in-between the controller platform and the controller application, which conforms to the OpenFlow policy. The SDN-aim Dos attack can be identified with FloodDefender with the utilization of new frequency features with the detection module. Evaluating the prototype with the implementation of FloodDefender, show the defense framework can identify and efficiently mitigate the Dos attack with little overhead. 
 
The work in \cite{64} demonstrates the dynamic provision of data connectivity with automatic failure recovery, at both control and data plane. Open standards and interfaces are used in demonstrating data provisioning, whereas the recovery part consist of ONOS at the control plane and logically centralize coordinated instances for robustness. The authors show how different devices can remain under control in the event of failure of any instances, that rely on the ATOMIX framework with dynamic real-time negotiation of devices. Ivan et al. \cite{143} present a comprehensive analysis on the security features of SDN/NFV that describe the strategies enable to monitor, protect and react to IoT security threats. In this, the author’s adaptation of SDN/NFV protection approached comparing it with conventional countermeasures. Tooska et al.\cite{144} identify the various potential attack scenario possible, such as denial of service and saturation attacks which is taken into consideration in the proposal design of control plane with OpenFlow. The work in \cite{81} summarizes the fundamental security challenges and problems that appear in the back-door of future mobile networks with several security issues and solutions are suggested experimentally with ONOS controller, which indicate much benefits. Danda and Swetha \cite{145}, present various security threats that can be resolved through SDN and the threat that can arise as a result of its implementation with the existence of smart programmable devices. Dmytro et al. \cite{146} present a comprehensive overview of related issues to security in software-defined networks and network virtualization functions. 

Khan et al. \cite{147}, evaluate security-related measures from both security and privacy perspective with comprehensive detail on the core and enabling technologies used in building 5G security model. Network softwarization security and physical layer security monitoring and management in 5G network and explore the security threat. Based on the availability of the literature review with provisioning of understanding the landscape security-related issues. Cao et al. \cite{148}, present an overview of the network architecture with security functionality of 3GPP 5G networks with a focus on new features and techniques that support IoT devices, Device-to-Device communication (D2D), vehicle to everything (V2X) communication and network slicing. The authors further present details in security features, security requirements, and existing security solutions by providing the current 5G network with security architecture provided by 3GPP. The authors have further shown the improvement in security aspect with smooth implementation and deployment of 5G when their work is put into consideration.

Rupprecht et al. \cite{149}, unify the security knowledge in mobile phone networks through a method that categories the known attacks. The authors apply the method to the existing attacks and defenses in all the previous network generations (i.e., 2G, 3G, and 4G) and identify the root causes of the attacks. This has given the opportunity in proposing defense mechanisms against such attacks in the current 5G mobile network, such as the problem of unsecured pre-authentication of traffic and jamming attacks that are present in the last three mobile network generations. The authors also show how to eliminate the class of downgrading attacks. Therefore, providing more security to the user's privacy. In \cite{150}, present and future security challenges in wireless mobile networks were analyzed and security vulnerabilities existing in the new technology adopted by 5G, as a result, the integration of new services which are absent in previous mobile network, such as IoT, smart home, connected vehicles and so on. This poses great security challenges and threats considering the technology to be used in 5G have their own security threats (i.e., MEC, SDN, and NFV). Solutions to the security threats and challenges are provided by the authors in knowing the risks pose and taking them into account from the initial design phase to the deployment phases, which will minimize the potential risk. 

The joint effort in \cite{151} propose a systematic analytical method for 5G Non-access stratum signaling security based on formal analysis that verifies 10 new 5G protocol vulnerabilities, with an improvement to Public Key Infrastructure (PKI) security mechanism to eliminate the vulnerabilities. In addition, the work of \cite{152}, further discuss the concept of agile security within 5G infrastructure with a focus on MEC and NFV, and how it can be embedded within the daily activities of mobile network operator with risk management as the key elements of agile security framework. Man-In-The-Middle (MITM) attack is implemented in the work of Pingle et al. \cite{153}, for instructional use in academic teaching cyber-security. Using the open-source Ettercap tool in kali Linux environment which is a sniffing tool and can be used in performing other attacks, such as DDOS, packet filtering, and Domain Name System (DNS) spoofing. In table \ref{tab.VIII}, we summarize the security-related publications in the context of open-source-defined wireless networks. HUANG et al. \cite{027} present the state-of-the-art solutions in security and privacy for V2X communication with security detailed cryptography-based scheme and trust-based scheme. However, 5G technology aims to provide more possibilities to V2X based communication which shows a great advantage. In this, the authors also present security architecture solutions for cellular-based communication with open challenges. 

\begin{table*}[t]
	\caption{Summary of important security publications in the context of open-source-defined wireless network}
	\label{tab.VIII}
	\centering
		\begin{tabular}{|c|c|c|c|}
			\hline
			\textbf{Area of Focus}             & \textbf{Main contribution}                                                                                                                                                                             & \textbf{Ref.} & \textbf{Relevant Challenges}                                                                                                                                                                                   \\ \hline
			Software-defined security          & \begin{tabular}[c]{@{}c@{}}The proposal of software defined \\ security (SDS) that is open and universal \\ and provide security services and security \\ management to security devices.\end{tabular} &          \cite{134}     & \begin{tabular}[c]{@{}c@{}}Analyzing the different types of \\ attacks and how the defender can \\ disable the network attack through \\ the realization of virtual security \\ functions (VSFs).\end{tabular} \\ \hline
			Security testbed                   & \begin{tabular}[c]{@{}c@{}}The proposal of security testbed for the \\ next-generation mobile networks that \\ enable researcher to perform various \\ security analysis.\end{tabular}                 &     \cite{136}         & \begin{tabular}[c]{@{}c@{}}Conducting relevant cyber-attack and \\ control measure on how the defense \\ measures are performed.\end{tabular}                                                                  \\ \hline
			Control and data plane             & \begin{tabular}[c]{@{}c@{}}The use of ONOS at the control plane \\ in logically centralizing coordinated \\ instances for achieving robustness.\end{tabular}                                           &   \cite{a133}           & \begin{tabular}[c]{@{}c@{}}The different devices have to remain \\ under the control of the ONOS in the \\ event of failure of any instances.\end{tabular}                                                     \\ \hline
			IoT security threats               & \begin{tabular}[c]{@{}c@{}}The introduction of a unique taxonomy \\ that discuss IoT vulnerabilities and \\ the amalgamation of IoT research trend.\end{tabular}                                       &        \cite{137}       & \begin{tabular}[c]{@{}c@{}}The severity of the IoT problem that\\ aid in the mitigation task.\end{tabular}                                                                                                     \\ \hline
			Control plane and OpenFlow         & \begin{tabular}[c]{@{}c@{}}The implementation of OpenFlow \\ communication channel with different \\ statistical properties.\end{tabular}                                                              &      \cite{138}         & \begin{tabular}[c]{@{}c@{}}The effective detection of saturation \\ attack in SDN.\end{tabular}                                                                                                                \\ \hline
			SDN/NFV DoS/DDoS attack            & \begin{tabular}[c]{@{}c@{}}The proposed mitigation of DoS/DDoS\\  with the cooperation of advanced \\ mechanism for a better reliable \\ cellular communication network.\end{tabular}                  & \cite{139}             & \begin{tabular}[c]{@{}c@{}}The limited deployment scenario \\ of the security mechanism of \\ 5G network.\end{tabular}                                                                                         \\ \hline
			RISV-V Processor attack            & \begin{tabular}[c]{@{}c@{}}Present a technique to prevent RISC-V \\ based embedded computing against \\ fault attack.\end{tabular}                                                                     &    \cite{140}           & \begin{tabular}[c]{@{}c@{}}Protecting the RISV-V processor \\ against fault injection \\ attack on the microchip.\end{tabular}                                                                                 \\ \hline
			Core network attack classification & \begin{tabular}[c]{@{}c@{}}Classification of cyber-attacks on \\ the core network, from the perspective \\ of a network protocol in the core network\end{tabular}                                      &       \cite{141}        & \begin{tabular}[c]{@{}c@{}}The creation of a new access path, \\ security downgrading and limitation \\ in security visibility.\end{tabular}                                                                   \\ \hline
		\end{tabular}
\end{table*} 
\subsection{Industry View Point on Security of Open 5G}
In this subsection, we discuss some of the viewpoints on open 5G security from the industry perspective. Across the industry domain, the potential transformation of 5G in businesses can be realized via various use cases. Ericsson in their white paper listed five core 5G security properties to ensure trustworthiness which include resilience, communication security, identity management, privacy, and security assurance. Ericsson's white paper focuses on the automatic recovery mechanism, security and privacy aspect of network slicing life cycle management, and further adjustment to SBA in support of various new cases with development in virtualization technology \cite{154E}. In another Erisson white paper, cloud-based open RAN deployment threats, vulnerabilities, and security are examined as well as vector attacks. Supply chain attacks, cross-container intrusions, and weak authentication are exploited to comprise application and data breaches. Confidentiality and integrity to realize strong cloud security-based life-cycle management and zero trust architecture with trust chain, and mutual authentication \cite{154D}. AT\&T published a similar concern in their white paper detailing security services to support 5G strategy including next-generation firewall, secure remote access solution, and sure web gateway to protect critical information from attack \cite{154F}. 

However, Nokia's white paper on 5G security discusses the 5G security assets to be protected, threats, risks, and mitigation steps in minimizing those risks. The adaptation of a global security assurance scheme-based Network Element Security Assurance Scheme (NESAS) for GSMA on mobile network security assurance methodology (SECAM) form 3GPP standards \cite{154G}. In Huawei 5G security white paper detail the security risk faced by 5G architecture, technology, and services in addressing the security challenges, based on 4G security, the 5G security can be achieved through the joint effort of all industries working together to share responsibilities and established a secure system via open and transparent 5G security ecosystem \cite{154H}.
\subsection{Current and Future Evolution}
In this subsection, we focus on current and future evolution in this new paradigm. Considering the tremendous and unlimited growth of data traffic from different devices, the utilization of next-generation RAN, SDN/NFV technology, and the integration of massive MIMO, and full-duplex in communication. All this requires an open-source platform which researchers from both academia and industry can use, as a real-world analysis of all these required elements. Open5GCore \cite{154}, is a new 5G stand-alone component and independent from the previous 4G EPC functionality. Open5GCore can enable a fast 5G innovation, demonstration and implementation of new concepts, and use cases with realistic evaluation. The Open5GCore release5 integrates 5G NR Stand-Alone (SA) off-the-shelf LTE and non-3GPP access networks, such as WiFi and 60 Ghz WiFi, that can run on a common hardware platform with container or VM deployment on a large number of virtualization environment. This is highly customizable, enabling instance deployment and addressing the need for specific use cases. 

In the work of Ullah et al. \cite{011}, propose an open-source framework based on Named Data Networking (NDN) technology and Edge Cloud Computing (ECC) in empowering the future internet, whose main target goal is achieving low latency and high bandwidth. The authors integrated the two technologies (i.e., NDN and ECC) to accomplish flexibility, privacy, and efficiency in IoT devices with various traffic loads. The implementation of NDN-based ECC framework with evaluated results obtained through testbeds and simulations shows NDN integration with ECC achieved latency and backbone network traffic, with a capacity of real-time result delivery to users and fast-enormous amount of data processed. 
Parvez et. al \cite{x015} surveyed the emerging technology to achieved low latency into three perspectives: 1) RAN; 2) CN; and 3) caching. To achieve low latency in the RAN domain, short packets frame modulation and coding schemes, transmission technique, cloud-RAN, the enforcement of QoS, and Quality of Experience (QoE) are studied.  However, the architecture of SDN, NFV, MEC, and the backhaul data traffic on the CN, and cache placement, content delivery for low latency in content download for caching are reviewed as well. A two-level MAC scheduling framework is proposed in \cite{z016} to handle the uplink and downlink transmission in network slice with different requirements across a shared RAN to achieve low latency with a dynamic radio resource management in meeting the required latency. In an effort to demonstrate the design flexibility of the two-level MAC scheduling framework, slice management decisions and its capability in enhancing data delivery to achieved uRLLC requirements are implemented.

Furthermore, the recent mobile layer-based heterogeneous network called moving network from the research community has proven to improve QoS for commuters boarding public transport. Due to fast-moving public transport pressure high amount of load on the CN, as a result of large group handovers. The above mention challenge motivates the work in \cite{x014} to surveyed moving networks, and mobile cells in filling the current literature gap and address future applications. Mobile cells can reduce the number of handovers by decoupling the CN from the boarded vehicle users, through the introduction of in-vehicle access link and backhaul access link to connect the out-of-vehicle traffic to the CN. The potential use cases with value-added applications of moving networks to the future cellular architecture, and challenges face revealed the author’s future research work.
In the work of Chen et al., \cite{155}, OpenFunction is proposed which is an extensible data plane abstraction protocol that is platform-independent of Software-defined Middle-boxes (SDM), with the core decision of actions/states/event operation to be defined within a uniform pattern. The authors implemented SDM consisting of OpenFunction controller-based Netmap, DPDK, and FPGA to verify the OpenFunction abstraction capability. 

The work in \citen{156}, introduces an open-source cellular network into MEC, which is defined by open-source software on general-purpose hardware based decouple MEC function and resources via NFV. The authors show a use case of open-source MEC scheme through demonstration in a test network. The analysis in \cite{73}, lead to the proposal of UOS in a similar fashion the work in \cite{157} analyze the security weakness in the control plane with provisioning framework focusing on the essential network properties required by the network. Open5G is proposed in \cite{158}, based OpenFlow (OF) and OF-Config, which is commonly used protocol in SDN wired and data centers. Multi-RAT RAN can be controlled through an open interface with flexibility and simplicity to the network operation with Open5G. The authors proposed Open5G due to the fragmented nature of the control plane in the multi-RAT RAN, lacking a global view of the network resources with hindering optimization of allocated resources. T-DCORAL \cite{159} is proposed to mitigate issues in Software-defined Data Center Networks (SD-DCNs) based Elastic Control Plane (ECP).
 
T-DCORAL can accelerate the control plane through the allocation of virtual CPU (vCPU) from a pool of vCPU to controllers in run time, on the other hand, the existing EPC resizes the controller pool. Results show T-DCORAL minimizes the control plane latency from 46 ms to 38 ms, likewise the average CPU load by 22\%, and the rule installation time by 64.28\%. SoftBox \cite{160}, a novel cellular CN architecture based SDN/NFV is proposed to achieve greater flexibility, efficiency, and scalability compared to the present day cellular core network, that implements a network policy required for processing of individual UE's data with signaling traffic to a light-weight, in-network and UE's traffic steering via it agent. SoftBox optimization efficiency drops the different types of data and control plane loads in the basic SoftBox by 51\%.

Next G alliance is a north american wireless technology advancement for 6G and beyond through a private sector initiative to poster full life-cycle research and development in manufacturing and standardization \cite{160A}. Hexa-X is a European 6G project with the vision to connect human, physical, and the digital world with 6G key enablers to include sub-THz transceiver technology, radio-based imaging, improve radio performance, and AI/ML, based on key areas such as connecting intelligence, and global service coverage \cite{160B}. Furthermore, the 6G KOM project R\&D, is a German new baseband architectural concept for 6G with the goals of novel ideas investigation to test and analyze D-band modules with respect to mobile communication \cite{160C}.   
\section{Open-Source Hardware}\label{GG}
In this section, we discuss open-source hardware which is classified into three categories: Baseband and Radio Frequency (RF), common hardware (x86, ARM), and open-source hardware (RISC-V). Hardware processors are required in running the operating system of a computer with the primary architecture of most processors to be Reduced Instructions Set Computer (RISC).
\subsection{Baseband and RF} 
Wireless communication systems are continuously growing to satisfy the diverse number of required wireless devices and the technicality designed of the RF system. However, integrated RF transceivers can reduce the design complexity of analog RF systems. The transceivers serve as an interface for easier integration with ASIC or FGPA in the implementation of baseband processing, which enable access to user data in the digital domain. Efforts in research to demonstrate experiment with new ideas to enhance wireless communication system starts with computer simulation in the first step. In addition, testing platforms are required in transferring the computer simulation into implementation validation. Moreover, SDR and SoC provide flexible research experimentation with wireless algorithm, whereas current SDR does not fully implement FGPA in SoC for the back-end receiver operation. This has motivated the work in \cite{x009} to demonstrate the hardware and automatic synthesizing, and implementation of baseband receiver-based wireless algorithm signal processor for FGPA, which uses SDR and SoC platform to present a real-world 6 GHz wireless experiment. The authors have taken into account the mitigation of the channel effect, and synchronization effect with respect to the baseband processing platform.  

The integration of digital baseband processor and ASIC to provide the wireless interface with Ultra-High-Frequency (UHF) Radio-Frequency-IDentification (RFID), based reader and tag for bio-signal healthcare monitoring system is presented in \cite{x012}. This ensures the hand-shake communication process between reader and tag. To achieve bi-directional communication the design process of reader and tag implement baseband processor ASICs on-chip and on FGPA respectively, and embedded in a wearable device to accomplish wireless communication with commercial RF. To demonstrate the simultaneous signal transmission of both RF and high-speed baseband through the implementation of superimposing electrical method, across graded-index silica multi-mode fiber to achieve 28 GHz RF and 28 Gbit/s pulse amplitude modulation of level-4 in the work of Ohtsuki et. al \cite{x013}. The scalability of the method describes the simultaneous transmission of dual-channel and effective utilization of transmitters under prescribed wavelength from 850 nm to 1550 nm across 50 m multi-mode fiber. 

The increasing momentum of billions of IoT devices has pushed researchers and manufacturers in rethinking the redesign of IoT devices, that are energy-efficient, scalable, accurate, and have authentication mechanisms at the devices level rather than the software level. This is because present authentication applications are not directly applicable to the IoT devices and are energy-consuming, based cryptography algorithm protocol. 
To identify the unique radio hardware impairment inflicted through the physical layer I/Q (i.e., in-phase and quadrature-phase) radio circuitry, from a pool of devices by deep-learning technique, Sankhe et al. \cite{x010} propose to optimize radio classification through a convolutional neural network (ORACLE) in solving the problem. ORACLE defined a new system-based Convolutional Neural Network (CNN). This technique specifies a particular radio feature-based hardware-centric in the transmitter chain. However, CNN accuracy of 80-95\% with implementation of COTS WiFi device and USRP radio in a static environment. The CNN technique can be scaled up through Impairment Hopping Spread Spectrum (IHOP) which is more resilient to spoofing attacks in classifying the number of radio by means of random binary sequence key.

Baseband complex signals (i.e., amplitude and phase) can be coded into pulse width and pulse position with binary signal to implement an efficient digital RF transmitter, through the process of RF modulation based uniform sampling on Pulse Width Modulation (PWM). RF-PWM carrier frequency serves as equal to the PWM first harmonic frequency such that the latter (i.e., PWM first harmonic), can serve as a carrier signal to obtain a comprehensive detailed model of uniform sampling RF-PWM that include in-band and side-band signal distortion. However, this distorted signal can be analyzed and discrete-time implementation of the model can be predicted to serve as characterization and simulation properties of fully RF digital transmitter \cite{x011}.    

\subsection{Common Hardware (x86, ARM)}

In today's computing trend, smartphones and tablets are shaping the landscape of computing, such that energy and power consumption are the constraints now to be considered. With this trend, ARM which is a RISC Instruction Set Architecture (ISA) and dominates the mobile market, whereas x86 is a Complex Instruction Set Computing (CISC) ISA and dominated the desktop/laptop, as well as server domain \, cite{161}. Both ARM and x86 are designed to optimize the different levels of performance with energy efficiency. With x86 there has been a long history of virtualization which is accomplished by executing multiple operating systems on the hardware processor, enabling encapsulation and use in the data centers. ARM virtualization extension can be used with a Linux-based virtual machine (KVM) in executing a particular software on ARMv8. ARM performance monitor extension can be used in accurate instruction counting, which requires the integration of processor model within the virtual platform (VP), which are the enabling technology in SoC (i.e., VP) \cite{165}. Digital information processing, privacy, and confidentiality is a concern due to cache attack that pose a great challenge.

In order to provide a solution to the challenge for cache attack on ARM, Flush + Time is proposed in \cite{166} which provides high accuracy and high-resolution flash-based cache attack, to solve the problem of flush cache lines and how precise timing is achieved. Experimental analysis shows the Flush + Time increase accuracy from 95\% to 99.3\%. ARM and x86 support open-source Architecture Code Analyzer (OSACA), which is a static analysis tool to execute and predict loop sequential time. This includes detection of loop dependencies, that turns into a versatile cross-architecture modeling tool \cite{167}. ARM-based server processor promotes the use of massive parallel High-Performance Computing (HPC) systems. This allows porting HPC applications to the architecture with a collection of recorded information about its performance, challenges, and benefits \cite{168}.
\subsection{Open-Source Hardware (RISC-V)} 
In this subsection, we discuss the RISC-V which is a popular architecture with simple command instructions that handle access to memory with load-stored commands. This runs faster due to the simplicity of the hardware instruction set. As such, a RISC-based processor can be designed faster because it contains a few chip-set \cite{160}. RISC demand currently surged due to the realization of IoT and used in embedded systems. The description of the open-source RISC-V design processor core targeted for near-threshold operation with the introduction of instructions, extension in optimizing the increase of computational density, and minimizing the pressure in shared memory can be analyzed in \cite{161}. RISC-V processor was developed in 2010 at Berkeley university as open-source instruction that lead the way for open-source hardware under Berkeley Software Division (BSD) license \cite{162}. RISC-V hardware architecture can provide a complete open-source computing system with the required freedom that is needed, e.g., toward an open-source operating system development project. 

RISC-V ISA free and modular, encourage the implementation of Intellectual Property (IP) cores that make it simple to use because of its open-source nature with an additional advantage compared to commercial CPU cores. However, the authors focus on existing open-source RISC-V CPU IP cores with three cores selected and evaluated \cite{163}. The FPGA-based SoC open-source CPU IP cores that support RISC-V design can accelerate the reused of IPs, with significant emphasis given to the design, test, verified, and shared in modularity level in order to increase productivity. Taiga \cite{164}, 32 bits RISC-V soft processor architecture can support RISC-V multiple operations with extension design support for Linux-based shared memory, which is highly configurable with a standardized interface. iFLEX an open-source high-density FPGA hardware co-processor enable
large-scale research and commercial application in the field of AI with diverse extraction of multimedia content through vector similarity searching with a high volume of data set entries\cite{037}. 

The open-source hardware accelerator designed for vector similarity searching (i.e., iFLEX) is capable of yielding 33.6 GB/s bandwidth with seamless reconfiguration designed-custom compute node board, based array of 21 FPGA to implement an arbitrary distance calculation. This is achieved through the implementation of distributed distance calculation with novel logic and software.
However, in support of open innovation and research, the authors provide the designed circuit board hardware, FPGA logic, and the host software. 
OpenNoC \cite{034} is an open-source infrastructure-based FPGA hardware accelerator, on lightweight deflection torus based Network-on-Chip (NoC) with Peripheral Component Interconnect express (PCIe) communication controller. Open-source (NoC) implementation fastens and saves the development time between hardware accelerators on FPGA, contrary to the traditional NoC that suffers from low clock performance when implemented with FPGA. 

In the work of \cite{035}, a modular cluster-based multi-core architecture that is flexible and reusable tailored in the implementation of different multi-core taxonomy is presented.
NoC is couple with multiple-processing clusters for communication that permit high-level designed capability in building many-core architecture-based-FPGA accelerators. The multi-core architecture consists of RISC-V Process Element (PE) couple to a bus-based interconnection, which serves intra-cluster communication through shared memory, with each PE posting a single RISC-V core tightly couple to a localized memory. For every research work, an expected achievable goal is expected and the proposed multi-architectural core obtained a scalable computing performance with a memory bandwidth of 4.3 Gb/s and 9 clusters utilizing SoC with 7.7\% power consumption. Heradio et al. \cite{033} presented a systematic mapping study that covered the literature review of open-source hardware in education, by classifying hundreds of publications according to: 1) guidance of published material; 2) the academic information about the use of open-source hardware; 3) and direction for the future research. The authors were motivated by the numerous publications proposing clear and distinct measures in scaling approach for effective learning purpose, that will promote the scientific concept, and student's creativity with new innovation.   
\section{Testbeds for Open-Source-Defined Wireless Network} \label{HH}
In this section, we discuss testbeds for open-source-defined wireless networks which are important platforms for research experimentation and prototyping. These testbeds provide the essential tools used by researchers and academia to perform tests that validate the components to be used (e.g., UE, eNodeB, EPC, etc.) with the performance for a novice technology in a wireless network. This is achieved on the foundation of general-purpose hardware (e.g., x86 and ARM), with decoupled network functions running as VNFs. However, we evaluate the test network to show the implementation of PoC framework design, that can use any of the open-source testbeds to implement and build mobile network functions.
Among these is the Mosaic5G which is the world's first ecosystem that is an open-source platform for 5G R\&D, which range from centralized network to mobile edge deployment which is complementary to OAI and leverage SDN/NFV. Mosaic5G provides applications to the edge and support services that are closer to the device with open-source APIs in fostering a third-party network developer. The APIs leverage the benefited values in deploying cellular network infrastructure with the control plane on the server-side to provide an on-demand resource monitoring of the network status. This open-source technology has changed the mobile network ecosystem by simplifying the operation of mobile network as a service and converging the mobile network into the cloud and edge technology with a click/plug and play software fashion for both RAN and CN. In addition, Mosaic5G offers to virtualize resources at the cloud and core for orchestration and split the control/user plane base service-based oriented architecture \cite{7}.

Furthermore, OAI is an open-source that offers a software implementation of all components ranging from UE, eNodeB, and core network which is achieved through the software radio front end connected to a host computer and enable innovation in the field of mobile network communication. Software-defined and virtualized wireless testbed-based OSS on COTs is presented in\cite{171}.
Security testbeds are essential in analyzing the challenge in security, privacy, and conducting cyber-security analysis on various network. Open-source project (i.e., MANO, ONOS, OpenStack) security testbed is established in \cite{136} which leverage these existing open-source projects in realizing a flexible network architecture.

TeFENS testbed provides a platform to experiment and demonstrate security services on real-time data with programmable OpenFlow enable switch \cite{135}. 
5GIIK is a novel testbed that offers the implementation, management, and orchestration of different network slices over the network domain with different access technology determine by different open-source tools implementation. 5GIIK provides features and additional capabilities which include real-time monitoring of both VNFs assigned to different VIMs and VMs \cite{172}. SimuLTE (System-level simulation) is a prominent model used in related research of 4G and 5G cellular communication. Based on OAI, cellular testbed is set up which measure the packet delay and inter-arrival time of low and high load scenario.

SimuLTE is an open-source platform integrated with OMNet++ and NET framework written in C++ and fully customizable with simple interface \cite{173}. OpenTestbed can be a significant tool for evaluating and bench-marking IoT solutions which consist of dedicated switches, servers, and a reservation with experimental management back-end which features all necessary tools in building a testbed from COTS. OpenTestbed is a complete open-source with open-source hardware project that has been adopted by various institutions \cite{125}. IoT platforms include TLSensing (i.e., an environment use in the monitoring of critical and non-critical parameters) and NetwOrked Smart Object (i.e., NOS). This integrate a real smart testbed, that enables fast deployment of IPv6 across TSCH (Time Slot Channel Hopping) mode of IEEE 802.15.4e, which reveals the data collected from sensors via a web interface.

COMOS testbed is city-scale advanced wireless tested design to support uRLLC and ECC, base software-defined radio node wireless communication. Cloud enhances Open Software-defined Mobile (COSMOS) wireless testbed enables researchers to experiment remotely on COSMOS testbed with programmable advantage to validate real-world and next-generation technology and application. This remote testbed deployed in New York City was built on the integration of mmWave, optical transport network, control and management software, as well as core and edge cloud \citen{y173}. In addition, 5TONIC is an open research platform for 5G related technology for research and innovation to both academia and industries. This provides 5G experimentation equipped resources to run and test complex open-source projects \citen{z173}. Fed4Fire+ is a Next Generation Internet federated testbed. This includes both wired, wireless, 5G, IoT, OpenFlow, cloud, and big data. Fed4Fire+ implement Open API that enables the interconnection of these different testbeds \citen{w173}.
\section{Lessons Learned}
In this section, we discuss lessons learned and future research direction with respect to open-source-defined wireless networks. Open-source-defined wireless networking frameworks, open-source software, open-source project, and testbeds, as well as security challenges, will be discussed.  
\subsection{Open-Source Software} 
\subsubsection{Lesson Learned}
We explore the historic evolution of open-source software and the potential solutions it provides from the terminals to the cloud infrastructure and operating system. We also highlighted the current trend of open-source software momentum extending the landscape of software solutions. We provide related research articles that are of benefit to our readers in choosing the appropriate software solutions to avoid any pitfalls to the user.  In fact, the problems open-source software faces are detailed to the readers to learn and fix errors and bugs which increases resiliency.

Although there are research-related articles and projects with open-source software most of which are related to the IoT realization to promote innovation. However, the distribution of programming languages, contributor’s specialization as well as the adopted dependencies can be considered for future research. 
\subsection{Open-Source Project}
\subsubsection{Lesson Learned}
A number of open-source projects to be exploited at both industrial and academic levels were discussed. The surging data traffic and compute power consumption challenges, for example, were addressed by OCP, creating open-source hardware with open specifications to ensure scalability, innovation, and energy-efficient hardware. Furthermore, the realization of SDN/NFV deployment challenges faced by integrating open-source projects was identified to our readers to avoid any pitfalls.  
and the challenges faced by open-source project, such as deployment, performance, and reliability.
\subsection{Frameworks}
\subsubsection{Lesson Learned}
The emergence of plethora of open-source frameworks has led both academia and industry to focus on the specifications to facilitate open implementation. This has recently pave the way for many research articles implementation of the framework with AI/ML technique [71], [72], [76]. In fact, global open-source community successive efforts promote multi-vendor products and innovations from the core network, access, edge, and terminal. The successive trend of categories of open-source-defined wireless networking frameworks from black-box, gray-box, and white-box network detail implementation, enable service providers to deploy infrastructure at low cost and vendor lock-in prevention [103].
To the best of our knowledge open-source frameworks domain consists of novel frameworks which required more research efforts from the community in promoting vendor agnostic products and innovation. However, there is still the need to integrate AI/ML technique for intelligent management. 
\subsection{Core Network}
\subsubsection{Lesson Learned}
The introduction of SBA enable telecom operators to flexible deployed virtual mobile network function on commodity servers with full control and route policy management that is centralize via the concept of SBA. This is a solution in providing software-defined, programmable, and distributed management in shifting away from communication protocol-centric. The MultiFire-based of LTE-like services in the unlicensed spectrum enabled business opportunities and extends the LTE protocol stack to implement flexible resource allocation and power consumption control, as well as enhancing eMBB and IoT applications [150], [151].
The contribution of MultiFire standalone LTE deployment in the 5 GHz unlicensed spectrum in the enhancement of eMBB and IoT applications and also in decreasing the latency as well as increasing the reliability of radio communication in unlicensed spectrum can be considered future research direction. 
\subsection{RAN}
\subsubsection{Lesson Learned}
The successive trend in achieving open-source-defined RAN is discuss in detail, from OpenBTS which provide more than one carrier grade at a time, to SDR that made the implementation of Open-source RAN possible on commodity server. However, experimental framework research is provided, e.g., IEEE 802.11 p based SDR OFDM to the readers to learned and explore the novel ideas (e.g., from the reference [173]).
As stated in reference [151] the challenges due to the LBT and other regulatory requirement from the introduction of LTE operations in unlicensed spectrum.  For example, MulteFire introduce the new uplink control channel designed and HARQ feedback mechanism. However, research efforts can be focus on solution implementation that enable 5G use cases in unlicensed and shared spectrum, as well as low latency communication.
\subsection{Security}
\subsubsection{Lesson Learned}
The wide exploitation of open-source in critical infrastructure provide a large surface of security attacks and sensitive consumer data increase the security requirements. Security can be accomplished at different level of the network, e.g., in the reference [212] experimentation with ONOS is conducted for mobile network. The deployment of security at different level as been well studied and provide hints to our readers on experimentation and precaution to avoid confidential data breached, e.g., in the references [211], [213], [214], [215], [221]-[233].
As our security review and analysis show, 5G use cases security architecture research need more focus on the IoT applications due to its diverse applications in our daily life activities as it provides a great surface of security risk.
\subsection{Testbed}
\subsubsection{Lesson Learned}
We have evaluated test network to show to the readers the implementation of PoC frameworks designed that can be exploited to use any of the open-source testbeds in the implementation of mobile network functions [22].
The diverse important platforms for research and prototyping of novel research innovation and ideas, e.g., the Mosaic5G provide a platform to conduct future R\&D. As presented in [215], security testbeds are essential in conducting future research based open-source projects in realization of flexible network architecture. 
\section{Open Research Issue and Future Research Direction}
In this section, we discuss open-source-defined wireless networks integration with some of the novel technologies, such as blockchain and AI/ML. 
However, now the question remains how does blockchain technology applied in telecommunication enable decentralization, anonymity, auditability, security, flexibility, efficiency and deployment cost. 
\subsection{Core Network and Blockchain}
Blockchain recently has gone beyond bitcoin (i.e. digital concurrency) to dominate the landscape of every technology, from telecommunication, IoT, cloud computing, to even our daily live activities. Blockchain technology which is known to be a distributed ledger database and enable information update securely with every node having all required information and the metadata. Each block in the blockchain enable independent storage unit which stores transaction related data. Taking into consideration the CN SBA feature to support flexibility, efficiency and openness. However, the technology enablers of 5G raise several issues such as transparency, decentralization, and reliability, because of the heterogeneous devices to be connected by 5G exposing grater surface to security attacks on confidential and data integrity. Blockchain can serve as a solution in bridging the gap to ensure transparency, trustworthiness, immutability, and data reliability in a plethora of new applications, therefore establishing trust among untrusted services.  

Blockchain sharding is proposed in \cite{w174} to enable horizontal scaling of partitioned several transactions and process independently. This involve building a number of clusters with multiple nodes (i.e., miners) in processing every shard in parallel. However, to verify data queries and authenticate its source they further proposed peer-to-peer oracle network, the blockchain oracle interface (trusted third party service provider) which verifies data authenticity. The fact that blockchain cannot access network external data. The structural incompatibility of blockchain design is address in \cite{w175}. The authors propose a novel temporal blockchain (conte) design base on federated byzantine fault tolerance. Conte deployed as cloud-native application (i.e., a micro-services inside docker containers). 
\subsection{RAN and Blockchain}
The heterogeneous and highly complexity of the RAN forces mobile networks operators (MNOs) to rely on their infrastructure and spectrum for data delivery to their subscribers, resulting to inefficiency and redundancy. The current state-of-the-art literature on levering blockchain in networks most currents works have focus on IoT, cloud and edge computing, and wireless sensor networks. Limited research article can be found on the blockchain integration in wireless communication. However, recent research works mostly have pay attention to integration of blockchain with RAN, for service latency, efficiency, flexibility, and security. The reference in \cite{w176} proposed the B-RAN to established trust between distinct groups for deep integration and efficient interaction by codification of six hierarchical functional layers in building trust with key enabling technologies in overcoming obstacles. Roaming network architecture based blockchain smart contract 5G networks is proposed in \cite{w177} to improves the visibility for mobile network operator’s activities in the visited networks, to enable fast payment and prevent fraudulent transactions. The importance of hash access in blockchain RAN is emphasize from a game theory perspective on transmission success probability, access delay, and network throughput \cite{w178}. Through the establishment of analytical model in evaluating the performance of B-RAN, hash access optimization is conducted for network throughput with simulation results to validate the proposed models.
\section{Conclusion}\label{II}

The open-source-defined wireless networks has created a huge impact in the ecosystem of wireless communication, in line with standards that create interoperability, openness, reduce cost, and create a healthy business ecosystem. This has become the game-changer in the recent 5G/6G mobile network.
However, the analysis made in this paper with respect to open-source projects that could benefit operators in recent 5G deployment and wireless networks is the infrastructure (i.e., programmable data plane), and management and orchestration (i.e., automation, orchestration, and analytics). Both open-source communities and standardization promote one another. The role of one (i.e., open-source community or standardization) can dominate the other. For example, Linux open-source community dominates over standardization in the operating system. In addition, 3GPP dominates over open-source community in wireless communication. The cohesive existence between the two will play a vital role that enables their coexistence in the wireless communication field. 

Furthermore, 3G and 4G suffered from a lack of interoperability, whereas 5G/6G was able to be free from this, due to its architectural design which provides openness via open-source projects. With open-source projects, great momentum is achieved in delivering services and products that are open, flexible, interoperable, and multi-vendor. Potential research directions are discussed, which include the integration of this paradigm into the next-generation wireless networks, security challenges, and privacy in open-source-defined wireless networks is also discussed. The trend surge of data traffic on infrastructure with key solutions is also provided to our readers. Finally, this comprehensive survey can serve as a guide for academia and industry in diving further into the context of open-source-defined wireless networks.  

\end{document}